\documentclass[final,numbers,1p,times]{elsarticle}
\usepackage{amsfonts,amssymb}
\usepackage{amsmath,amscd}
\usepackage{amsmath}
\usepackage{graphicx}
\usepackage{epstopdf}
\newcommand\dif{\mathrm{d}}
\usepackage{subfigure}

\usepackage[plainpages=false, pdfpagelabels]{hyperref}
\usepackage{mathrsfs}

\hypersetup{
    linktocpage=true,
    colorlinks=true,
    bookmarks=true,
    citecolor=blue,
    urlcolor=red,
    linkcolor=Maroon,
    citebordercolor={1 0 0},
    urlbordercolor={1 0 0},
    linkbordercolor={.7 .8 .8},
    pdfpagelabels=true,
    }

\begin{document}
\makeatletter
\def\ps@pprintTitle{%
  \let\@oddhead\@empty
  \let\@evenhead\@empty
  \let\@oddfoot\@empty
  \let\@evenfoot\@oddfoot
}
\makeatother
\begin{frontmatter}

\title{A Network-Based Meta-Population Approach to  Model Rift Valley fever Epidemics}

\author[1]{Ling Xue}
 \ead{lxue@ksu.edu}
 \author[2]{H. M. Scott}
 \ead{hmscott@vet.k-state.edu}
 \author[3]{Lee W. Cohnstaedt}
 \ead{Lee.Cohnstaedt@ars.usda.gov}
\author[1]{Caterina Scoglio\corref{a}}
 \cortext[a]{Corresponding Author}

 \ead{caterina@k-state.edu}
 \address[1]{Department of Electrical \&  Computer Engineering,\\ Kansas State University, \ U.S. \ 66506}
\address[2]{Department of Diagnostic Medicine/Pathobiology, \\ Kansas State University, \ U.S. \ 66506}
\address[3]{ Center for Grain and Animal Health Research, \\ United States Department of Agriculture, \ U.S. \ 66502}

\begin{abstract}
Rift Valley fever virus (RVFV) has been expanding its geographical distribution with important implications for both human and animal health. The emergence of Rift Valley fever (RVF) in the Middle East, and its continuing presence in many areas of Africa, has negatively impacted both medical and veterinary infrastructures and human morbidity, mortality, and economic endpoints. Furthermore, worldwide attention should be directed towards the broader infection dynamics of RVFV, because suitable host, vector and environmental conditions for additional epidemics likely exist on other continents; including Asia, Europe and the Americas. We propose a new compartmentalized model of RVF and the related ordinary differential equations  to assess disease spread in both time and space; with the latter driven as a function of contact networks. Humans and livestock hosts and two species of vector mosquitoes are included in the model. The model is based on weighted contact networks, where nodes of the networks represent geographical regions and the weights represent the level of contact between regional pairings for each set of species. The inclusion of  human, animal, and vector movements among regions is new to RVF modeling. The movement of the infected individuals is not only treated as a possibility, but also an actuality that can be incorporated into the model. We have tested, calibrated, and evaluated the model using data from the recent $2010$ RVF outbreak in South Africa as a case study; mapping the epidemic spread within and among three South African provinces. An extensive set of simulation results shows the potential of the proposed approach for accurately modeling the RVF spreading process in additional regions of the world. The benefits of the proposed model are twofold: not only can the model differentiate the maximum number of infected individuals among different provinces, but also it can reproduce the different starting times of the outbreak in multiple locations. Finally, the exact value of the reproduction number is numerically computed and upper and lower bounds for the reproduction number are analytically derived in the case of homogeneous populations.

\end{abstract}


\end{frontmatter}

\section{Introduction}

Rift Valley fever (RVF) is a viral zoonosis with enormous health and economic impacts on domestic animals and humans \cite{ Linthicum2007}, in countries where the disease is endemic and in others where sporadic epidemics and epizootics have occurred.  An outbreak in South Africa in $1951$ was estimated to have infected $20,000$ people and killed $100,000$ sheep and cattle \cite{DepartmentEnvironment2010, Sellers1982}. In Egypt in $1977$, there were $18,000$ human cases with 698 deaths resulting from the disease \cite{DepartmentEnvironment2010, Sellers1982}. While RVF is endemic in Africa, it also represents a threat to Europe and Western hemispheres \cite{Chevalier2005, Gaff2007}. In $1997-1998$ Kenya experienced the largest recorded outbreak with $89,000$ human cases and $478$ death. The first recorded outbreak outside of Africa occurred in the Arabian  peninsula in $2000-2001$ and caused $683$ human cases and $95$ deaths \cite{FloridaDepartmentofHealthRVF}. Tanzania and Somalia reported $1000$ human cases and $300$ deaths from an outbreak that was associated with above-normal rainfall in the region in $2006-2007$ \cite{FloridaDepartmentofHealthRVF}.
  Rift Valley fever virus (RVFV) is generally distributed through regions of Eastern and Southern Africa where sheep and cattle  are present \cite{Woods2002}. It can cause morbidity (ranging from nondescript fever to meningo-encephalitis and hemorrhagic disease) and mortality (with case fatality rates of $0.2-5$\%$)$ in humans \cite{ Linthicum2007}. The main economic losses of RVF in livestock arise due to abortion and mortality, which tends to be higher in young animals \cite{Clements2007, Woods2002}, and bans on livestock exports during an epidemic \cite{Clements2007, Anyamba2001}.

Rift Valley fever virus was first isolated from the blood of a newborn lamb in $1931$ and later from the blood of adult sheep and cattle \cite{WHORVF2010, Anyamba2001}. Domestic ruminants and humans are among the mammalian hosts demonstrated to amplify RVFV \cite {Kasari2008} and \it Aedes \rm and \it Culex \rm are believed to be the main arthropod vectors \cite{Chevalier2005} .  Rift Valley fever virus can be transferred vertically from females to their eggs in some  species of the \it Aedes \rm mosquitoes \cite{Gaff2007, Linthicum1985}. The disease has been shown to be endemic in semi-arid zones, such as northern Senegal \cite{Zeller1997, Chevalier2005, Martin2008}, and RVF epidemics often appears at $5-15$ year cycles \cite{Martin2008}.  As noted earlier, RVFV has already spread outside Africa, to Yemen and Saudi Arabia \cite{Chevalier2005, CDC2007}.  The species of vectors that are capable of transmitting RVFV have a wide global distribution \cite{Gubler2002} and there is therefore a distinct possibility for the virus to spread out of its currently expanding geographic range \cite{Clements2007}. A pathways analysis \cite {Kasari2008} has shown that the RVF virus might be introduced into the United States in several different ways \cite {Kasari2008, Konrad2010} and that analysis identified several regions of the United States that are most susceptible to RVFV introduction. It is therefore desirable to develop effective models to better understand the potential dynamics of RVF in heretofore unaffected regions and then develop efficient mitigation strategies in case this virus appears in the Western hemisphere \cite{Gaff2007}. Such preparedness can help avoid a rapid spread of the virus throughout North America, as happened with the West Nile virus during the last decade \cite{Chevalier2005, Gaff2007}.

A RVF disease risk mapping model was developed by \cite{Anyamba2009}. The authors observed sea surface temperature (SST) patterns, cloud cover, rainfall, and ecological indicators (primarily vegetation) via satellite data to evaluate different aspects of climate variability and their relationships to disease outbreaks in Africa and the Middle East \cite{Anyamba2002, Anyamba2006}. The researchers successfully predicted areas where outbreaks of RVF in humans and animals were expected using climate data for the Horn of Africa from December $2006$ to May $2007$.  An ordinary differential equation (ODE) mathematical model was developed by \cite{Gaff2007}. The model is both an individual-based and deterministic model. The authors analyzed the stability of the model and tested the importance of the model parameters. However, neither human population parameters nor spatial (or, network) aspects are explicitly incorporated in the model. Another theoretical mathematical model on RVFV dynamic transmission was proposed \cite{ Mpeshe2011}. This model is also an individual based model. The most important parameters to initial disease transmission and the endemic equilibrium  have been carried out.

In this paper, we present a novel model incorporating \it Aedes \rm and \it Culex \rm mosquito vector, and livestock and human host populations. Our model is based on weighted contact networks, where nodes of the networks represent geographical regions and weights represent the level of contact between regional pairs for each vector or host species. Environmental factors such as rainfall, temperature, wind and evaporation are incorporated into the model. For each subpopulation, a set of ordinary differential equations describes the dynamics of the population  in a specific geographical location, and the transitions among the different compartments, after contracting the virus. We compute the lower and upper bounds of the reproduction number for homogeneous populations, explain their biological meaning, and numerically compare the bounds with exact values.

We test, calibrate, and evaluate the model using the recent $2010$ RVF outbreak in South Africa as a case study, mapping the epidemic spread in three South African provinces: Free State, Northern Cape, and Eastern Cape. An extensive set of simulation results shows the potential of the proposed approach to accurately describe the spatial–temporal  evolution of RVF epidemics.

The paper is organized as follows: $1)$ in Section \ref {section:model}, we describe our compartmentalized mathematical model, present the lower bound and upper bound of the reproduction number for homogeneous populations. $2)$ in Section \ref {section:case study}, we introduce the case study using outbreak data from South Africa, $2010$, $3)$   in  Section \ref {section:conclusionanddiscussion}, we conclude our work. In the Appendix, we show how we derive the  bounds for  the reproduction number for homogeneous populations.

\section{Compartmentalized Mathematical Model}\label{section:model}
We have constructed Compartmentalized Mathematical Models based on the principle of RVFV transmission. The parameters used in the model are shown in Table \ref{table:parameters}.
\subsection {Homogeneous Populations Model}
\begin{figure}[h]
  \centering
   \includegraphics[angle=0,width=14.5cm,height=14cm]{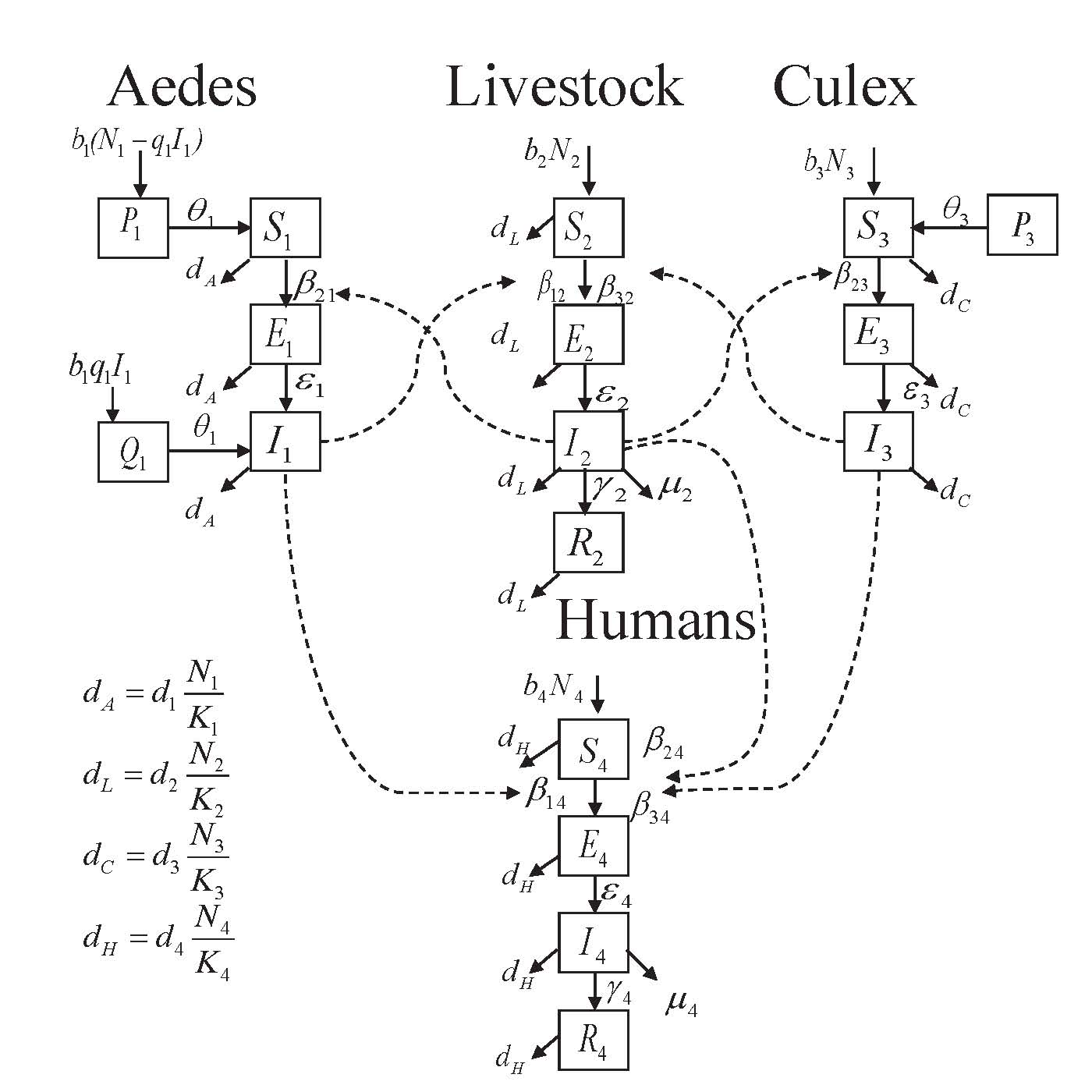}
 \caption{Flow diagram of RVFV transmission with each species, namely, \it Aedes \rm mosquitoes, \it Culex \rm mosquitoes, livestock, and humans homogeneously mixed (the solid lines represent transition between compartments and the dash lines represent the transmission between different species)}
\label{fig:first}
\end{figure}

The principle of RVFV transmission between different species is shown in Figure \ref{fig:first}. All the Aedes 
 ssp. and Culex  spp.  we are going to discuss only include the mosquitoes that are competent vectors of Rift Valley fever. In this paper, the Culex parameters are based on Culex Tarsalis mosquitoes and Aedes parameters are based on Ae. dorsalis mosquitoes. The main vectors, \it Aedes \rm and \it Culex \rm  mosquitoes and the main hosts, livestock and humans are considered in the model.  We use an SEI compartmental model in which individuals are either in a susceptible (S) state, an exposed (E) state, or an infected state (I) for both \it Aedes \rm and \it Culex \rm mosquitoes, and an SEIR compartmental model in which individuals are either in a susceptible (S) state, an exposed (E) state, an infected state (I), or a recovered (R) state for both livestock and human populations.  Infectious \it Aedes \rm mosquitoes can not only transmit RVFV to susceptible livestock and humans but also to their own eggs \cite{Gaff2007, Linthicum1985}.   \it Culex \rm mosquitoes acquire the virus during blood meals on an infected animal and then amplify the transmission of RVFV through blood meals on livestock and humans \cite{WHORVF2010}.   Direct ruminant-to-human contact is the major (though not only) way for humans to acquire the infection \cite{Anyamba2009, Davies2006}. Accidental RVFV infections have been recorded in laboratory staff handling blood and tissue from infected animals \cite{Anyamba2009}. Usually, humans are thought of as dead end hosts that do not contribute significantly to propagation of the epidemic \cite{Chevalier2005}. There has been no direct human-to-human transmission of RVFV in field conditions recorded thus far \cite {Kasari2008}. The mosquitoes will not spontaneously recover once they become infectious \cite{Gaff2007}. Livestock and humans either perish from the infection or  recover \cite{Gaff2007}. All four species have a specified incubation period \cite{WHORVF2010}.  The model is based on a daily time step. \it Aedes \rm  and \it Culex \rm  mosquitoes
 are distributed among susceptible $S_{a}$, exposed $E_{a}$  and infected $I_{a}$ compartments. The subscript $a=1$ denotes \it Aedes \rm  and $a=3$ denotes \it Culex \rm. The size of each adult mosquito population  is  $N_{1}=S_{1}+E_{1}+I_{1}$ for adult \it Aedes \rm mosquitoes and $N_{3}=S_{3}+E_{3}+I_{3}$ for adult \it Culex \rm mosquitoes. The livestock and human hosts contain susceptible $S_{b}$, exposed $E_{b}$, infected $I_{b}$ and recovered $R_{b}$ individuals. The subscript $b=2$ and $b=4$ denote livestock and humans respectively. The size of host populations is  $N_{b}=S_{b}+E_{b}+I_{b}+R_{b}$.  The four populations are modeled with a specified carrying capacity  $K_1$,  $K_2 $, $K_3$,  $K_4 $ respectively.  \\

\begin{table}
\centering
\begin{tabular}{|p{35pt}|p{160pt}|p{80pt}|p{35pt}|p{55pt}|}
 \hline
Parameter & Description & Value& Units & Source\\
\hline
$\beta_{12}$  &  contact rate: \it Aedes \rm to livestock & $0.002$& $1/$day &\cite{Gaff2007}\\
\hline
$\beta_{21}$ &contact rate: livestock to \it Aedes \rm  & $0.0021$ &$1/$day &\cite{Gaff2007}\\
\hline
$\beta_{23}$ & contact rate: livestock to \it Culex \rm  &$0.000003$ &$1/$day &\cite{Gaff2007}\\
\hline
$\beta_{32}$ & contact rate: \it Culex \rm to livestock  &$ 0.00001$  & $1/$day & \cite{Gaff2007}\\
\hline
$\beta_{14}$ &  contact rate: \it Aedes \rm to humans   & $0.000046$   &$1/$day & Assume\\
\hline
$\beta_{24}$&  contact rate: livestock to humans    &$0.00017$   &$1/$day &\cite{Mpeshe2011}\\
\hline
$\beta_{34}$&  contact rate: \it Culex \rm to humans      & $0.0000001$     &$1/$day &Assume\\
\hline
$\gamma_2$& recover rate in livestock                  &$0.14$  &$1/$day & \cite{Gaff2007}\\
\hline
$\gamma_4$ &recover rate in humans                              &$0.14$ &$1/$day & \cite{statspopulation2010, statstoursurvey2010, statsAgriculture2010}\\
\hline
$d_1$ &  death rate of \it Aedes \rm mosquitoes                        &$ 0.025$            &$1/$day &\cite{Gaff2007}\\
\hline
$d_2$ & death rate of livestock                                        &$1/3650$          &$1/$day & \cite{Gaff2007}\\
\hline
$d_3$ &death rate of \it Culex \rm mosquitoes                             &$0.025$       &$1/$day & \cite{Gaff2007}\\
\hline
$d_4$   & death rate of  humans                                        &$1/18615$  &$1/$day& \cite{statspopulation2010, statstoursurvey2010, statsAgriculture2010}\\
\hline
$b_1$ &number of \it Aedes \rm eggs laid per day                   &$0.05 $          &$1/$day & \cite{Gaff2007}\\
\hline
$b_2$ & daily birthrate of livestock                               & $0.0028$         &$1/$day & \cite{Gaff2007}\\
\hline
$b_3$ & number of \it Culex \rm eggs laid per day                   & weather dependent       &$1/$day & \cite{Gong2010}\\
\hline
$b_4$& daily birthrate of humans                                    &$1/14600$       &$1/$day & \cite{statspopulation2010, statstoursurvey2010, statsAgriculture2010}\\
\hline
$1/\epsilon_1$ &incubation period in \it Aedes \rm mosquitoes &$ 6$  &days      &\cite{Gaff2007}\\
\hline
$1/\epsilon_2$ &incubation period in livestock                &$4  $   &days      &\cite{Gaff2007}\\
\hline
$1/\epsilon_3$  &incubation period in \it Culex \rm  mosquitoes    &$6 $&days & \cite{Gaff2007}\\
\hline
$1/\epsilon_4$ &incubation period in humans                     & $4 $&days & \cite{WHORVF2010}\\
\hline
$\mu_2$ &mortality rate in livestock  & $0.0312$          &$1/$day &\cite{Gaff2007}\\
\hline
$\mu_4$&mortality rate in humans  &$0.0001    $           &$1/$day & \cite{statspopulation2010, statstoursurvey2010, statsAgriculture2010}\\
\hline
$q_1$ & transovarial transmission rate in \it Aedes \rm      & $0.05 $ &-  &\cite{Gaff2007}\\
\hline
$1/\theta_1$&development time of \it Aedes \rm  &$15    $          &days      &\cite{Gaff2007}\\
\hline
$\theta_3$&development rate of \it Culex \rm   &  weather dependent          &$1/$day      &\cite{Gong2010}\\
\hline
$K_1$ &carrying capacity of \it Aedes \rm mosquitoes             &$1000000000  $ &- &\cite{newton1992model}\\
\hline
$K_2$&carrying capacity of livestock                           &$ 10000000 $       &-& Assume\\
\hline
$K_3$& carrying capacity of \it Culex \rm mosquitoes&$1000000000 $  &- &\cite{newton1992model}\\
\hline
$K_4$ &carrying capacity of humans                              &$10000000$ &- &Assume\\
\hline
$f$&fraction of those working with animals& $0.82$ &- & \cite{NICD2010}\\
\hline
$\tau$&return rate& $3$ & times/day &\cite{Balcan2009}\\
\hline
$p$&reduction in $\omega^2_{ij}$ due to infection& $\frac{1}{2}$ & - &Assume\\
\hline
\end{tabular}
\caption{Parameters of the compartmentalized mathematical model}
\label{table:parameters}
\end{table}
 \subsubsection{\it Aedes \rm Mosquito Population Model}
\allowdisplaybreaks
\begin{align}
  \frac{\dif P_{1}}{\dif t} &= b_1 \left(N_{1}-q_1I_{1} \right) -\theta_1P_{1}\\
  \frac{\dif Q_{1}}{\dif t} &= b_1 q_1I_{1} -\theta_1Q_{1} \\
 \frac{\dif S_{1}}{\dif t} &=\theta_1P_{1}-\beta_{21}S_{1}I_{2}/N_{2}-d_1S_{1}N_{1}/K_1\\\
  \frac{\dif E_{1}}{\dif t} &=\beta_{21}S_{1}I_{2}/N_{2}-\varepsilon_1E_{1}-d_1E_{1}N_{1}/K_1\\\
 \frac{\dif I_{1}}{\dif t} &=\theta_1Q_{1}+\varepsilon_1E_{1}-d_1I_{1}N_{1}/K_1\\
  \frac{\dif N_{1}}{\dif t} &=\theta_1(P_{1}+Q_{1})-d_1N_{1}N_{1}/K_1
\end{align}

where:\\
$P_1$=the number of uninfected \it Aedes \rm mosquito eggs\\
$Q_1$=the number of infected \it Aedes \rm mosquito eggs \\
$S_1$=the number of susceptible \it Aedes \rm mosquitoes\\
$E_1$=the number of  exposed \it Aedes \rm mosquitoes\\
$I_1$=the number of infected \it Aedes \rm mosquitoes\\
$N_1$=the total number of  \it Aedes \rm mosquitoes\\

The above model is a modified SEI model with  compartments P and Q. Compartments P and  Q represent uninfected eggs and infected eggs  respectively. The total number of eggs laid each day is $ b_1N_{1}$ with $ b_1 q_1I_{1}$ infected eggs and  $b_1N_{1}-b_1 q_1I_{1}$ uninfected eggs \cite{Gaff2007}.  After development period, $\theta_1P_{1}$  develop into susceptible adult mosquitoes and $\theta_1Q_{1}$   develop into infected adult  mosquitoes \cite{Gaff2007}.  There are   $d_1X_{1}N_{1}/K_1$ mosquitoes removed from compartment $X$ due to natural death.  Compartment $X$ can be P, Q, S, E, and I here. The number  of \it Aedes \rm mosquitoes infected by livestock is denoted by $\beta_{21}S_{1}I_{2}/N_{2}$ which is proportional to the density of infected  \it Aedes \rm mosquitoes \cite{Gaff2007}.    After incubation period,  $\varepsilon_1E_{1}$ \it Aedes \rm mosquitoes transfer to infected compartment \cite{Gaff2007}.  \\

\allowdisplaybreaks
\subsubsection{\it Culex \rm Mosquito Population Model}

\begin{align}
  \frac{\dif P_{3}}{\dif t} &=b_3N_{3}-\theta_3P_{3}\\
 \frac{\dif S_{3}}{\dif t} &=\theta_3P_{3}-\beta_{23}S_{3}I_{2}/N_{2}-d_3S_{3}N_{3}/K_3\\
  \frac{\dif E_{3}}{\dif t} &=\beta_{23}S_{3}I_{2}/N_{2}-\varepsilon_3E_{3}-d_3E_{3}N_{3}/K_3\\
  \frac{\dif I_{3}}{\dif t} &=\varepsilon_3E_{3}-d_3I_{3}N_{3}/K_3\\
  \frac{\dif N_{3}}{\dif t} &=\theta_3P_{3}-d_3N_{3}N_{3}/K_3
\end{align}

where:\\
$P_3$=the number of uninfected \it Culex \rm mosquito eggs\\
$S_3$=the number of susceptible \it Culex \rm mosquitoes\\
$E_3$=the number of  exposed \it Culex \rm mosquitoes\\
$I_3$=the number of infected \it Culex \rm mosquitoes\\
$N_3$=the total number of  \it Culex \rm mosquitoes\\

Besides compartment $S$, $E$, $I$,  compartment P is added to represent  uninfected eggs. Only uninfected eggs are included because the female \it Culex \rm mosquitoes do not transmit RVFV vertically \cite{Gaff2007}. The total number of eggs laid each day is $ b_3N_{3}$. There are  $d_3X_{3}N_{3}/K_3$ \it Culex \rm mosquitoes removed due to natural death.  Compartment $X$ can be P,  S, E, and I here. After development period, $\theta_3P_{3}$  eggs develop into susceptible adult \it Culex \rm mosquitoes and become secondary vectors \cite{Gaff2007}.  The number of infected \it Culex \rm mosquitoes from contact with livestock is denoted by $\beta_{23}S_{3}I_{2}/N_{2}$ which is proportional to the percentage of infected livestock \cite{Gaff2007}. After incubation period, $\varepsilon_3E_{3}$  \it Culex \rm mosquitoes transfer from exposed compartment to infected compartment \cite{Gaff2007}. \\

 \subsubsection{Livestock  Population Model}

\allowdisplaybreaks
\begin{align}
  \frac{\dif S_{2}}{\dif t} &=b_2N_{2}-d_2S_{2}N_{2}/K_{2}-\beta_{12}S_{2}I_{1}/N_{1}
-\beta_{32}S_{2}I_{3}/N_{3} \\
  \frac{\dif E_{2}}{\dif t} &=\beta_{12}S_{2}I_{1}/N_{1}
+\beta_{32}S_{2}I_{3}/N_{3}-\varepsilon_2E_{2}-d_2E_{2}N_{2}/K_{2}\\
  \frac{\dif I_{2}}{\dif t} &=\varepsilon_2E_{2}-\gamma_2I_{2}-\mu_2I_{2}-d_2I_{2}N_{2}/K_{2}\\
\frac{\dif R_{2}}{\dif t} &=\gamma_2I_{2}-d_2R_{2}N_{2}/K_{2}\\
  \frac{\dif N_{2}}{\dif t} &=b_2N_{2}-d_{2}N_{2}N_{2}/K_{2}-\mu_2I_{2}\\[-5pt]\nonumber
\end{align}

where:\\
$S_2$=the number of susceptible livestock\\
$E_2$=the number of  exposed livestock\\
$I_2$=the number of infected livestock\\
$N_2$=the total number of  livestock\\

 There are $ b_2 N_{2}$  livestock born,  $d_2X_{2}N_{2}/K_2$ livestock removed due to natural death \cite{Gaff2007},  and $\mu_2I_{2}$ livestock dying of the infection each day \cite{Gaff2007}.  Compartment $X$ can be S, E, I, and R here. Following incubation period,  $\varepsilon_2E_{2}$  livestock transfer from exposed compartment to infected compartment \cite{Gaff2007}. The  number of livestock infected by   \it Aedes \rm mosquitoes and  \it Culex \rm mosquitoes are denoted as  $\beta_{12}S_{2}I_{1}/N_{1}$
and $\beta_{32}S_{2}I_{3}/N_{3}$ respectively \cite{Gaff2007}.  Following infection period, $\gamma_2I_{2}$ livestock recover from RVFV infection \cite{Gaff2007}.   \\
\subsubsection{Human Population Model}
\allowdisplaybreaks
\begin{align}
  \frac{\dif S_{4}}{\dif t} &=b_4N_{4}
-\beta_{14}S_{4}I_{1}/N_{1}-f\beta_{24}S_{4}I_{2}/N_{2}-
\beta_{34}S_{4}I_{3}/N_{3}-d_4S_{4}N_{4}/K_{4}\\
 \frac{\dif E_{4}}{\dif t} &=\beta_{14}S_{4}I_{1}/N_{1}+f\beta_{24}S_{4}I_{2}/N_{2}+
\beta_{34}S_{4}I_{3}/N_{3}-d_4E_{4}N_{4}/K_{4}-\varepsilon_4E_{4}\\
  \frac{\dif I_{4}}{\dif t} &=\varepsilon_4E_{4}-\gamma_4I_{4}-\mu_4I_{4}-d_4I_{4}N_{4}/K_{4}\\
 \frac{\dif R_{4}}{\dif t} &=\gamma_4I_{4}-d_4R_{4}N_{4}/K_{4}\\
  \frac{\dif N_{4}}{\dif t}&=b_4N_{4}-d_4N_4N_{4}/K_{4}-\mu_4I_{4}
\end{align}

where:\\
$S_4$=the number of susceptible humans\\
$E_4$=the number of  exposed humans\\
$I_4$=the number of infected humans\\
$N_4$=the total number of  humans\\

There are $ b_4 N_{4}$  humans born,  $d_4X_4N_{4}/K_4$ humans removed from compartment $X$ due to natural death,  and $\mu_4I_{4}$ humans dying of RVFV infection each day.   Compartment $X$ can be S, E, I, and R here.  The number of  humans that  acquire the infection from \it Aedes \rm mosquitoes, \it Culex \rm mosquitoes, and livestock is $\beta_{14}S_{4}I_{1}/N_{1}$, $\beta_{34}S_{4}I_{3}/N_{3}$, and  $f\beta_{24}S_{4}I_{2}/N_{2}$ respectively. We assume only those who work with animals can be infected by animals. Therefore, a coefficient $f\ ( 0<f<1)$ which represents the fraction of humans working with animals is multiplied by $\beta_{24}S_{4}I_{2}/N_{2}$. After incubation period, $\varepsilon_4E_{4}$ humans transfer to infected compartment and $\gamma_4I_{4}$ humans transfer to recovered compartment after infection period.\\
\subsubsection{Environmental  Parameters for \it Culex \rm}
The equation $(\ref{equation:developmentrate})$  is used to model the development rate of \it Culex \rm mosquitoes \cite{Gong2010}. The daily egg laying rate expressed in equation (\ref {equation:birthrate}) is a function of moisture \cite{Gong2010}. Moisture in equation $(\ref{equation:Moisture})$ is obtained by summing the difference of precipitation \cite{NCDC2010} and evaporation (mm) \cite{Linacre1977} over the proceeding $7$ days \cite{Gong2010}. In the equations $(\ref{equation:developmentrate}) $ to $(\ref{equation:evaporation})$, $A$, $ HA$, $HH$, $K$, $TH$, $E_{max}$, $ E_{var}$, $E_{mean}$, $b_0$ are parameters \cite{Gong2010} which are described in  Table \ref{table:culexparameters}. This model is specific for West
Nile virus model in $2010$ in the northern US. 
\allowdisplaybreaks

\begin{align}
 \theta_3(Temp,t)&=A*\frac{(Temp(t)+K)}{298.15}*\frac{exp[\frac{HA}{1.987}*(\frac{1}{298.15}-\frac{1}{Temp(t)+K})]}{1+exp[\frac{HH}{1.987}*(\frac{1}{TH}-\frac{1}{Temp(t)+K})]}\label{equation:developmentrate}\\
b_3 (Temp, precipitation, t)&=b_0+\frac{Emax}{1+exp[-\frac{Moisture(t)-Emean}{Evar}]}\label{equation:birthrate}\\
Moisture(t)&=\sum^t_{D=t-6}participation(D)-evaporation(D) \label{equation:Moisture}\\
Evaporation(t)&=\frac{700 (Temp(t)+0.006h)/(100-latitude)}{80-Temp(t)}\nonumber\\&+\frac{15(Temp(t)-T_d(t))}{80-Temp(t)}mm/day\label{equation:evaporation}
\end{align}
Where:\\
$Temp(t)=$air temperature in units of $^{o}C$ \cite{Linacre1977}\\
$latitude=$the latitude (degrees) \cite{Linacre1977}\\
$T_d(t)=$the mean dew-point  in units of $^{o}C$ \cite{Linacre1977}\\
$h=$the elevation (meters) \cite{Linacre1977}\\
$K=$ Kelvin parameter \cite{Linacre1977}\\

\begin{table}
\centering
\begin{tabular}{|p{35pt}|p{160pt}|p{40pt}|p{55pt}|}
 \hline
Parameter & Description & Value & Source\\
\hline
$A$&parameter of the development rate & $0.25$&\cite{Gong2010}\\
\hline
$HA$&parameter of the development rate& $28094$ &\cite{Gong2010}\\
\hline
$HH$&parameter of the development rate& $35692 $& \cite{Gong2010}\\
\hline
$TH$&parameter of the development rate & $298.6$ & \cite{Gong2010}\\
\hline
$b_0$&minimum constant fecundity rate & $3$ & \cite{Gong2010}\\
\hline
$E_{max}$&maximum daily egg laying rate & $3$ &\cite{Gong2010}\\
\hline
$E_{mean}$&value at which moisture index=$ 0.5 E_{max}$&$0$ &\cite{Gong2010}\\
\hline
$E_{var}$&  the variance of the daily egg laying rate  &$12$ &\cite{Gong2010}\\
\hline
\end{tabular}
\caption{Parameters of the model for \it Culex \rm}
\label{table:culexparameters}
\end{table}

\subsubsection{The Reproduction Number for Homogeneous Populations}

The reproduction number $R_0$ is defined as: `` The average number of secondary cases arising from an average primary case in an entirely susceptible population" \cite{diekmann2000mathematical}. The reproduction number is used to predict whether the epidemic will spread or die out. There are several methods used to compute $R_0$.  One of these methods computes the reproduction number as the spectral radius \cite{diekmann2000mathematical, heffernan2005perspectives} of the next generation matrix  \cite{diekmann2000mathematical, heffernan2005perspectives}.

The next generation matrix  is defined as $FV^{-1}$, and the matrices $F$ and $V$ are determined as:
$$F= [\frac{\partial \mathscr{F}_i (x_0)}{\partial x_j}], \           V= [\frac{\partial \mathscr{V}_i(x_0)}{\partial x_j}]$$

where $x_j$ is the number or proportion of infected individuals in compartment $j$, $j=1, 2, 3,... \  , m$, $m$ being the total number of infected compartments, $x_0$ is the disease free equilibrium vector, $\mathscr{F}_i $ is the rate of appearance of new infections in compartment $i$, and  $\mathscr{V}_i=\mathscr{V}_i^{-}-\mathscr{V}_i^{+}$ with $\mathscr{V}_i^{-}$ denoting the transfer of individuals out of compartment $i$ and $\mathscr{V}_i^{+}$ denoting the rate of transfer of individuals into compartment $i$ \cite{van2002reproduction}. The $(i, j)$ entry of $F$ represents the rate at which infected individuals in  compartment $j$ produce infected individuals in compartment $i$ \cite{van2002reproduction}. The $(j, k)$ entry of $V^{-1}$ represents the average time that an individual spends in compartment $j$, where $i, j, k=1, 2, 3,... \  , m$ \cite{van2002reproduction}. Finally, the  $(i, k)$ entry of $FV^{-1}$ represents the expected number of infected individuals in compartment $i$ produced by the infected individuals in compartment $k$ \cite{van2002reproduction}.\\

For our homogeneous population model, we found that

\begin{align}
R_0^H \leqslant R_0\leqslant R_0^H+q_1 \label{equation:R0boundd}
\end{align}
where\\

\begin{align}
R_0 ^{H}= \sqrt{
\frac{\varepsilon_2}{(b_2+\varepsilon_2)
(b_2+\gamma_2+\mu_2)}
\Big[\frac{\varepsilon_1\beta_{12}\beta_{21}}{b_1
 (b_1+\varepsilon_1)}
+\frac{\varepsilon_3\beta_{32}\beta_{23}}{b_3(b_3+\varepsilon_3)}
\Big] }\label{equation:R0H1}
\end{align}

See the Appendix for the derivation details, the biological interpretation, and the comparison among exact values and bounds for the reproduction number.\\

 \subsection {Meta-Population Model}
A meta-population model is a model with several subpopulations. It assumes homogeneity within each  subpopulation and heterogeneity among different subpopulations.
The \it Aedes \rm  and \it Culex \rm  mosquitoes in location $i\ ( i=1,2,3, \cdots, n)$,
 are distributed among susceptible $S_{ai}$, exposed $E_{ai}$  and infected $I_{ai}$ compartments. The subscript $a=1$ denotes \it Aedes \rm  and $a=3$ denotes \it Culex \rm. The size of each adult mosquito population in location $i$ is  $N_{1i}=S_{1i}+E_{1i}+I_{1i}$ for adult \it Aedes \rm mosquitoes and $N_{3i}=S_{3i}+E_{3i}+I_{3i}$ for adult \it Culex \rm mosquitoes. The livestock and human hosts contain susceptible $S_{bi}$, exposed $E_{bi}$, infected $I_{bi}$ and recovered $R_{bi}$ individuals. The subscript $b=2$ and $b=4$ denote the  livestock and humans, respectively. The size of host populations in location $i$ is  $N_{2i}=S_{2i}+E_{2i}+I_{2i}+R_{2i}$ for livestock hosts and $N_{4i}=S_{4i}+E_{4i}+I_{4i}+R_{4i}$ for human hosts.  The four populations are modeled with a specified carrying capacity  $K_1$,  $K_2 $, $K_3$,  $K_4 $ respectively.

\subsubsection{Movement between Nodes}
\begin{figure}[!htbp]
\centering
\subfigure[Mosquito movement network (Mosquitoes can move from node $ i$ to node $j_1$, $j_2$, and $j_3$ and vice versa due to wind. We  assume mosquitoes  do not return to the node they are from.)]{
\includegraphics[angle=0,width=6.1cm,height=5.9cm]{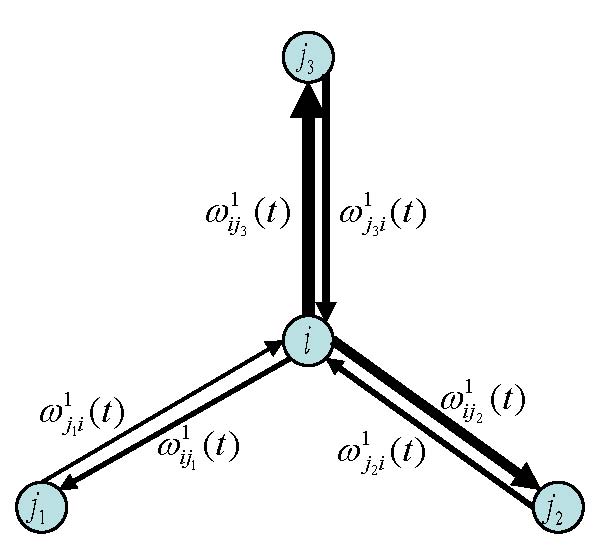}}
\hspace{0.1in}
\subfigure[Livestock movement network (Livestock can move from node $ i$ to node $j_1$, $j_2$, and $j_3$ and vice versa due to trade.  We  assume livestock  do not return to the node they are from.)]{
\includegraphics[angle=0,width=6.1cm,height=5.9cm]{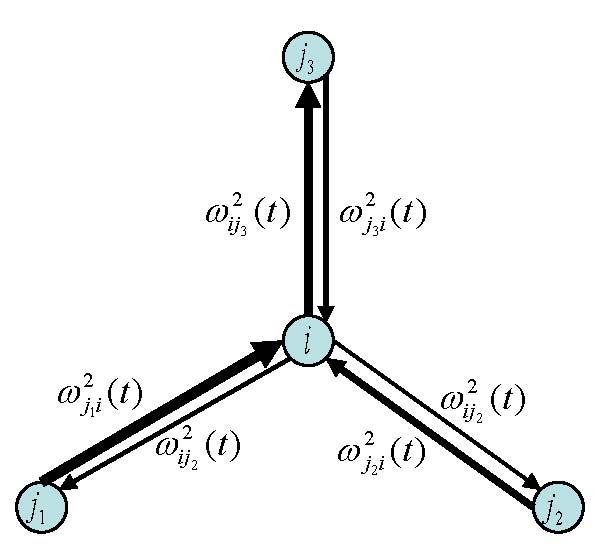}}
\hspace{0.1in}
\subfigure[ Human movement  network (Humans can commute from node $ i$ to node $j_1$, $j_2$, and $j_3$ and vice versa.  We  assume humans  return to the node they are from.)]{
\includegraphics[angle=0,width=6.1cm,height=5.9cm]{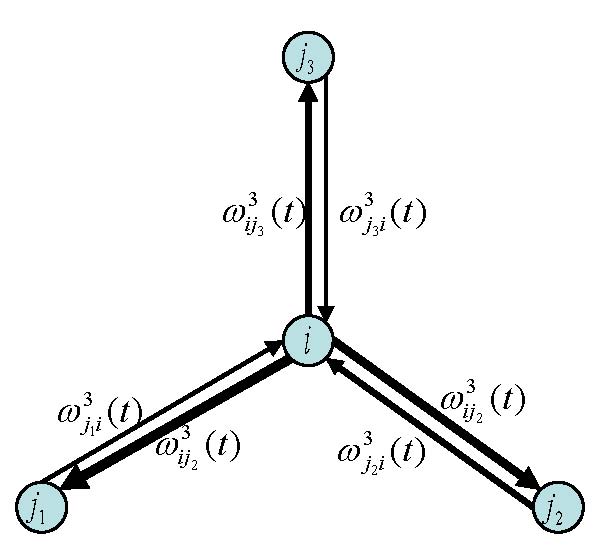}}
\hspace{0.1in}
\caption{Network graphs with node $i$ which has three neighbors as an example}
\label{fig:networkgraph}
\end{figure}

We used weighted networks for each compartment of the four species as is shown in Figure \ref{fig:networkgraph}. The superscripts of $\omega$ on the left hand side of  equations $(\ref{equation:mosquitoweight})$, (\ref{equation:animaloweight}), and $(\ref{equationhumanweight})$ represent the movement of different species. The number '$1$' in the superscript means the movement of \it Aedes \rm or \it Culex \rm population, '$2$' means the livestock movement, and '$3$' means the human movement in the superscript.
The subscript ${ij}$ of $\omega_{ij}$  means that the direction of the movement is from location $i$ to location $j$.
The difference in the thickness of the lines represent the difference in weight. Thicker lines represent the larger weight.
The weight for each population is between $0$ and $1$. RVFV has been documented to be spread by wind \cite{Sellers1982}. Wind dispersal of mosquitoes has changed geographic distribution and accelerated the spread of RVFV to new geographic areas \cite {Kasari2008}. Some locations can become secondary epidemic sites after the virus has been introduced (especially in irrigated areas, e.g. Gazeera in Sudan or rice valleys in the center of Madagascar) \cite{Martin2008}.
Livestock trade and transport also can affect the geographic distribution of RVF \cite{Chevalier2005}. One critical objective in developing effective models is to determine the major factors involved in the disease spreading process.
Therefore, we parameterize the weight due to mosquito movement with wind \cite {Kasari2008, Chowdhury2010},  livestock movement due to transportation to feedlots or trade centers \cite {swanson2001cattle}, and human mobility due to commuting  \cite{Balcan2009} as shown in equations $(\ref{equation:mosquitoweight})$, (\ref{equation:animaloweight}), and $(\ref{equationhumanweight})$, respectively.  The movement rate of infected livestock is reduced due to infection \cite{WHORVF2010}. We use  the wind data \cite{WeatherUnderground2010} in  Bloemfontein, which is the capital of Free State, as the wind of Free State Province, that of Kimberley, which is the capital of Northern Cape, as the wind of Northern Cape Province and that of Grahams town, which is the center of Eastern Cape Province, as the wind of  Eastern Cape Province.
The distance vector is calculated with longitude and latitude  in the center of each location. The number of animals sold \cite{statsAgriculture2010} and the number of livestock in the feedlots \cite{Olivier2004} are factors of weight for livestock movement. Distance, human population, commuting rate, and return rate \cite{statspopulation2010} affect the weight for human movement.
Weight for mosquito movement is decided by distance and the projection of wind in the direction of distance vector \cite{Chowdhury2010}.

\allowdisplaybreaks
\begin{align}
\omega^1_{ij}&=c_1\frac{\vec{W_{i}}\cdot \vec{D_{ij}}}{|\vec{D_{ij}}|}\frac{1}{|\vec{D_{ij}}|} \label{equation:mosquitoweight}\\
\omega^2_{ij}&=c_2\frac{FM_j}{FM_i}\frac{1}{|\vec{D_{ij}}|} \label{equation:animaloweight}\\
\sigma_{ij}&=c_3\frac{N^{\alpha}_{4i}N^\gamma_{4j}}{e^{\beta|\vec{D_{ij}}|}} \\
\omega^3_{ij}&=\frac{\sigma_{ij}}{N_{4i}} \label{equationhumanweight}\\
\omega_{i}&=\sum^n_{j=1, j \neq i}\omega^3_{ij}
\end{align}

Here:\\
$\vec{W_{i}} $= the  wind vector in location $i$ \cite{Chowdhury2010}\\
$\vec{D_{ij}} $ = the distance vector from location $i$ to location $j$\\
$\omega^1_{ij}(t)$ = the weight for mosquitoes moving from location $i$ to location $j$\\
$\omega^2_{ij}(t)$ = the weight for livestock moving from location $i$ to location $j$\\
$\sigma_{ij}(t)$ = the number of commuters between location $i$ and location $j$\\
$FM_i =$ the number of animals in markets and feedlots in location $i$\\

 \subsubsection{\it Aedes \rm Movement between Nodes}

\allowdisplaybreaks
\begin{align}
  \frac{\dif P_{1i}}{\dif t} &= b_1 \left(N_{1i}-q_1I_{1i} \right) -\theta_1P_{1i}\\
  \frac{\dif Q_{1i}}{\dif t} &= b_1 q_1I_{1i} -\theta_1Q_{1i} \\
 \frac{\dif S_{1i}}{\dif t} &=\theta_1P_{1i}-\beta_{21}S_{1i}I_{2i}/N_{2i}-d_1S_{1i}N_{1i}/K_1+\sum^n_{j=1, j \neq i}\omega^1_{ji}S_{1j}-\sum^n_{j=1, j \neq i}\omega^1_{ij}S_{1i}\\
  \frac{\dif E_{1i}}{\dif t} &=\beta_{21}S_{1i}I_{2i}/N_{2i}-\varepsilon_1E_{1i}-d_1E_{1i}N_{1i}/K_1+\sum^n_{j=1, j \neq i}\omega^1_{ji}E_{1j}-\sum^n_{j=1, j \neq i}\omega^1_{ij}E_{1i}\\
 \frac{\dif I_{1i}}{\dif t} &=\theta_1Q_{1i}+\varepsilon_1E_{1i}-d_1I_{1i}N_{1i}/K_1+\sum^n_{j=1, j \neq i}\omega^1_{ji}I_{1j}-\sum^n_{j=1, j \neq i}\omega^1_{ij}I_{1i}\\
  \frac{\dif N_{1i}}{\dif t} &=\theta_1(P_{1i}+Q_{1i})-d_1N_{1i}N_{1i}/K_1+\sum^n_{j=1, j \neq i}\omega^1_{ji}S_{1j}-\sum^n_{j=1, j \neq i}\omega^1_{ij}S_{1i}+\sum^n_{j=1, j \neq i}\omega^1_{ji}E_{1j} \nonumber \\&-\sum^n_{j=1, j \neq i}\omega^1_{ij}E_{1i}+\sum^n_{j=1, j \neq i}\omega^1_{ji}I_{1j}-\sum^n_{j=1, j \neq i}\omega^1_{ij}I_{1i}
\end{align}
 The change in the number of \it Aedes \rm mosquitoes due to mobility in  compartment $X$ is given as $\sum^n_{j=1, j \neq i}\omega^1_{ji}X_{1j}-\sum^n_{j=1, j \neq i}\omega^1_{ij}X_{1i}$ \cite{Keeling2008}.
\subsubsection{ \it Culex \rm Movement between Nodes}
\allowdisplaybreaks
\begin{align}
  \frac{\dif P_{3i}}{\dif t} &=b_3N_{3i}-\theta_3P_{3i}\\
 \frac{\dif S_{3i}}{\dif t} &=\theta_3P_{3i}-\beta_{23}S_{3i}I_{2i}/N_{2i}-d_3S_{3i}N_{3i}/K_3+\sum^n_{j=1, j \neq i}\omega^1_{ji}S_{3j}-\sum^n_{j=1, j \neq i}\omega^1_{ij}S_{3i}\\
  \frac{\dif E_{3i}}{\dif t} &=\beta_{23}S_{3i}I_{2i}/N_{2i}-\varepsilon_3E_{3i}-d_3E_{3i}N_{3i}/K_3+\sum^n_{j=1, j \neq i}\omega^1_{ji}E_{3j}-\sum^n_{j=1, j \neq i}\omega^1_{ij}E_{3i}\\
  \frac{\dif I_{3i}}{\dif t} &=\varepsilon_3E_{3i}-d_3I_{3i}N_{3i}/K_3+\sum^n_{j=1, j \neq i}\omega^1_{ji}I_{3j}-\sum^n_{j=1, j \neq i}\omega^1_{ij}I_{3i}\\
  \frac{\dif N_{3i}}{\dif t} &=\theta_3
P_{3i}-d_3N_{3i}N_{3i}/K_3+\sum^n_{j=1, j \neq i}\omega^1_{ji}S_{3j}-\sum^n_{j=1, j \neq i}\omega^1_{ij}S_{3i}+\sum^n_{j=1, j \neq i}\omega^1_{ji}E_{3j}\nonumber\\&-\sum^n_{j=1, j \neq i}\omega^1_{ij}E_{3i}+\sum^n_{j=1, j \neq i}\omega^1_{ji}I_{3j}-\sum^n_{j=1, j \neq i}\omega^1_{ij}I_{3i}
\end{align}
  The change in the number of  \it Culex \rm mosquitoes in compartment $X$ due to movement is given as $\sum^n_{j=1, j \neq i}\omega^1_{ji}X_{3j}-\sum^n_{j=1, j \neq i}\omega^1_{ij}X_{3i}$ \cite{Keeling2008}.
 \subsubsection{ Livestock Movement between Nodes}

\allowdisplaybreaks
\begin{align}
  \frac{\dif S_{2i}}{\dif t} &=b_2N_{2i}-\beta_{12}S_{2i}I_{1i}/N_{1i}-\beta_{32}S_{2i}I_{3i}/N_{3i} -d_2S_{2i}N_{2i}/K_{2}+\sum^n_{j=1, j \neq
i}\omega^2_{ji}S_{2j}-\sum^n_{j=1, j \neq
i}\omega^2_{ij}S_{2i}\nonumber\\&
\\
  \frac{\dif E_{2i}}{\dif t} &=\beta_{12}S_{2i}I_{1i}/N_{1i}
+\beta_{32}S_{2i}I_{3i}/N_{3i}-\varepsilon_2E_{2i}-d_2E_{2i}N_{2i}/K_{2}+\sum^n_{j=1, j \neq i}\omega^2_{ji}E_{2j}-\sum^n_{j=1, j \neq i}\omega^2_{ij}E_{2i}\\
  \frac{\dif I_{2i}}{\dif t} &=p\sum^n_{j=1, j \neq i}\omega^2_{ji}I_{2j}-p\sum^n_{j=1, j \neq i}\omega^2_{ij}I_{2i}-d_2I_{2i}N_{2i}/K_{2}+\varepsilon_2E_{2i}-\gamma_2I_{2i}-\mu_2I_{2i}\\
\frac{\dif R_{2i}}{\dif t} &=\sum^n_{j=1, j \neq i}\omega^2_{ji}R_{2j}-\sum^n_{j=1, j \neq i}\omega^2_{ij}R_{2i}+\gamma_2I_{2i}-d_2R_{2i}N_{2i}/K_{2}\\
  \frac{\dif N_{2i}}{\dif t} &=b_2N_{2i}-d_{2}N_{2i}N_{2i}/K_{2}-\mu_2I_{2i}+\sum^n_{j=1, j \neq
i}\omega^2_{ji}S_{2j}-\sum^n_{j=1, j \neq
i}\omega^2_{ij}S_{2i}+\sum^n_{j=1, j \neq i}\omega^2_{ji}E_{2j}\nonumber\\&-\sum^n_{j=1, j \neq i}\omega^2_{ij}E_{2i}+p\sum^n_{j=1, j \neq
i}\omega^2_{ji}I_{2j}-p\sum^n_{j=1, j \neq i}\omega^2_{ij}I_{2i}+\sum^n_{j=1, j \neq i}\omega^2_{ji}R_{2j}-\sum^n_{j=1, j \neq i}\omega^2_{ij}R_{2i}\\[-5pt]\nonumber
\end{align}

The change in the number of animals due to movement in susceptible, exposed, and recovered compartment is   $\sum^n_{j=1, j \neq i}\omega^2_{ji}X_{2j}-\sum^n_{j=1, j \neq i}\omega^2_{ij}X_{2i}$ \cite{Keeling2008}  for livestock. Concerning the animals in the infected compartment, we assume that the movement rate of the infected livestock is $p \ (0<p<1)$ of  livestock in other compartments. This value of the movement rate has been selected in the absence of further information.
\subsubsection{Human Movement between Nodes}

\allowdisplaybreaks
\begin{align}
  \frac{\dif S_{4i}}{\dif t} &=b_4N_{4i}-d_4S_{4i}N_{4i}/K_{4}
-\frac{\beta_{14}S_{4i}I_{1i}/N_{1i}}{1+\sigma_i/\tau}-\frac{\beta_{24}fS_{4i}I_{2i}/N_{2i}}{1+\sigma_i/\tau}-\frac{
\beta_{34}S_{4i}I_{3i}/N_{3i}}{1+\sigma_i/\tau}\nonumber\\&-\sum^n_{j=1, j \neq i}\frac{\beta_{14}S_{4i}I_{1j}/N_{1j}\sigma_{ij}/\tau}{1+\sigma_{i}/\tau}-\sum^n_{j=1, j \neq i}\frac{\beta_{24}fS_{4i}I_{2j}/N_{2j}\sigma_{ij}/\tau}{1+\sigma_{i}/\tau}-\sum^n_{j=1, j \neq i}\frac{
\beta_{34}S_{4i}I_{3j}/N_{3j}\sigma_{ij}/\tau}{1+\sigma_{i}/\tau}\\
 \frac{\dif E_{4i}}{\dif t} &=\frac{\beta_{14}S_{4i}I_{1i}/N_{1i}}{1+\sigma_i/\tau}+\frac{\beta_{24}fS_{4i}I_{2j}/N_{2i}}{1+\sigma_i/\tau}+\frac{
\beta_{34}S_{4i}I_{3i}/N_{3i}}{1+\sigma_i/\tau}+\sum^n_{j=1, j \neq i}\frac{\beta_{14}S_{4i}I_{1j}/N_{1j}\sigma_{ij}/\tau}{1+\sigma_{i}/\tau}\nonumber\\&+\sum^n_{j=1, j \neq i}\frac{\beta_{24}fS_{4i}I_{2j}/N_{2j}\sigma_{ij}/\tau}{1+\sigma_{ij}/\tau}+\sum^n_{j=1, j \neq i}\frac{
\beta_{34}S_{4i}I_{3j}/N_{3j}\sigma_{ij}/\tau}{1+\sigma_{i}/\tau}-d_4E_{4i}N_{4i}/K_{4}-\varepsilon_4E_{4i}\\
  \frac{\dif I_{4i}}{\dif t} &=\varepsilon_4E_{4i}-\gamma_4I_{4i}-\mu_4I_{4i}-d_4I_{4i}N_{4i}/K_{4}\\
 \frac{\dif R_{4i}}{\dif t} &=\gamma_4I_{4i}-d_4R_{4i}N_{4i}/K_{4}\\
  \frac{\dif N_{4i}}{\dif t}&=b_4N_{4i}-d_4N_{4i}N_{4i}/K_{4}-\mu_4I_{4i}
\end{align}
The humans from location $i$ can stay in  location $i$ or move to location $j$  at time $t$  \cite{Balcan2009}. The number of humans infected by \it Aedes \rm mosquitoes, \it Culex \rm mosquitoes and livestock is $\beta_{14}(S_{4ii}\frac{I_{1i}}{N_{1i}}+\sum^n_{j=1, j \neq i}S_{4ij}\frac{I_{1j}}{N_{1j}})$ \cite{Balcan2009}, $\beta_{34}(S_{4ii}\frac{I_{3i}}{N_{3i}}+\sum^n_{j=1, j \neq i}\beta_{34}S_{4ij}\frac{I_{3j}}{N_{3j}})$ \cite{Balcan2009}, and $f\beta_{24}(S_{4ii}\frac{I_{2i}}{N_{2i}}+\sum^n_{j=1, j \neq i}S_{4ij}\frac{I_{2j}}{N_{2j}})$ \cite{Balcan2009} respectively.\\

where:\\
$S_{4ii}$= the number of humans that are from location $i$ and stay in location $i$ at time $t$ \cite{Balcan2009}. \\
$S_{4ij}$= the number of humans that are from location $i$ and stay in location $j$ at time $t$ \cite{Balcan2009}.  \\
$\omega_{ij}^3$=  the commuting rate between subpopulation $i$ and each of its neighbor $j$  \cite{Balcan2009}\\
$\omega_{i}$ = daily total rate of commuting for population $i$ \cite{Balcan2009}\\

The change in the number of susceptible humans that are from location $i$ and stay in location $i$ is given  \cite{Balcan2009} by the following expression.
$$\frac{\partial S_{4ii}}{\partial t} =\sum^n_{j=1, j \neq i}\tau S_{4ij}-\sum^n_{j=1, j \neq i}\omega_{ij}^3S_{4ii}$$

The change in the number of susceptible humans that are from location $i$ and stay in location $j$ is given  \cite{Balcan2009} by the following expression.
$$\frac{\partial S_{4ij}}{\partial t} =\omega_{ij}^3S_{4ii}-\tau S_{4ij}$$

We can get the solution of  $S_{4ii}$ and  $S_{4ij}$ through the above two equations at the equilibrium.

\begin{align}
S_{4ii}&=\frac{S_{4i}}{1+\omega_{i}/\tau}\\
S_{4ij}&=\frac{S_{4i}}{1+\omega_{i}/\tau}\omega_{ij}^3/\tau
\end{align}

\section{Case Study: South Africa $2010$}\label{section:case study}
We have used data from the South African RVF outbreak in $2010$ as a case study.
\subsection{Incidence Data Analysis}
\begin{figure}[http!]
\centering
\subfigure[Simulation result and incidence data for sheep in Free State Province]{
\includegraphics[angle=0,width=6.5cm,height=5.2cm]{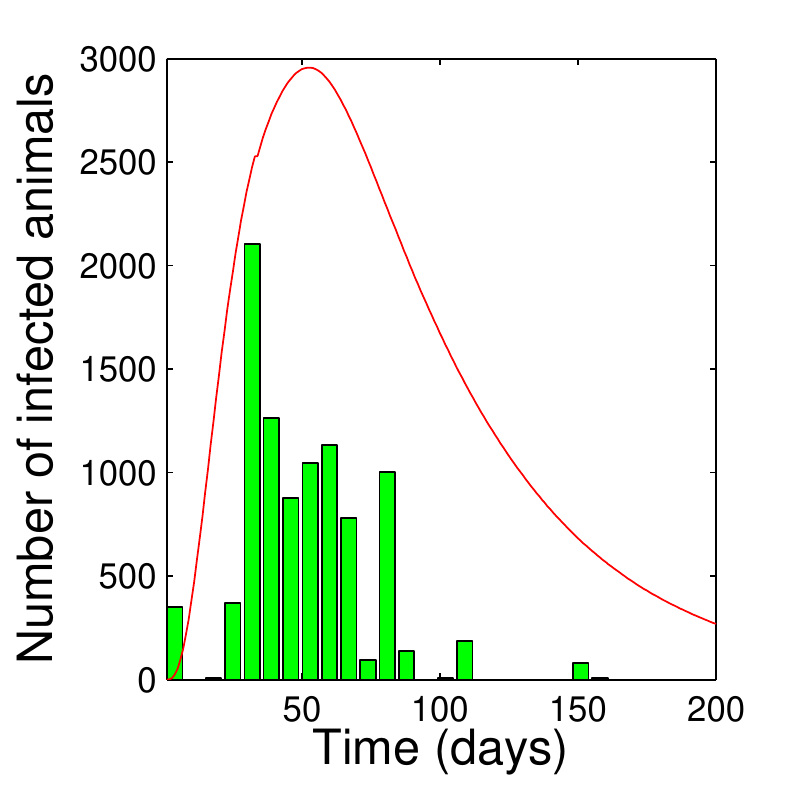}}
\hspace{0.0001in}
\subfigure[Simulation result and incidence data for humans in Free State Province]{
\includegraphics[angle=0,width=6.5cm,height=5.2cm]{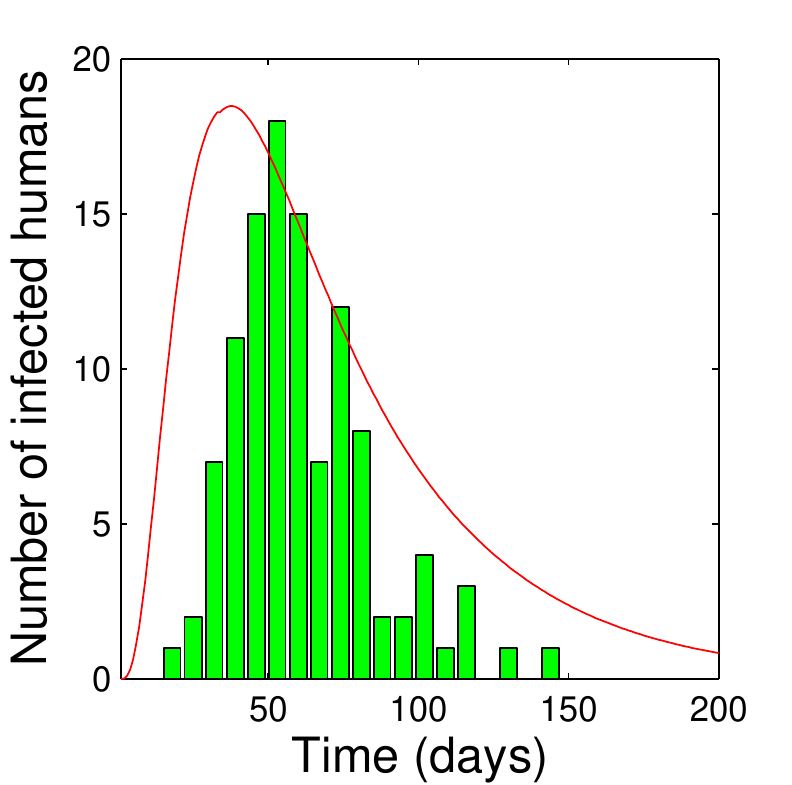}}
\subfigure[Simulation result and incidence data for sheep in Northern Cape Province]{
\includegraphics[angle=0,width=6.5cm,height=5.2cm]{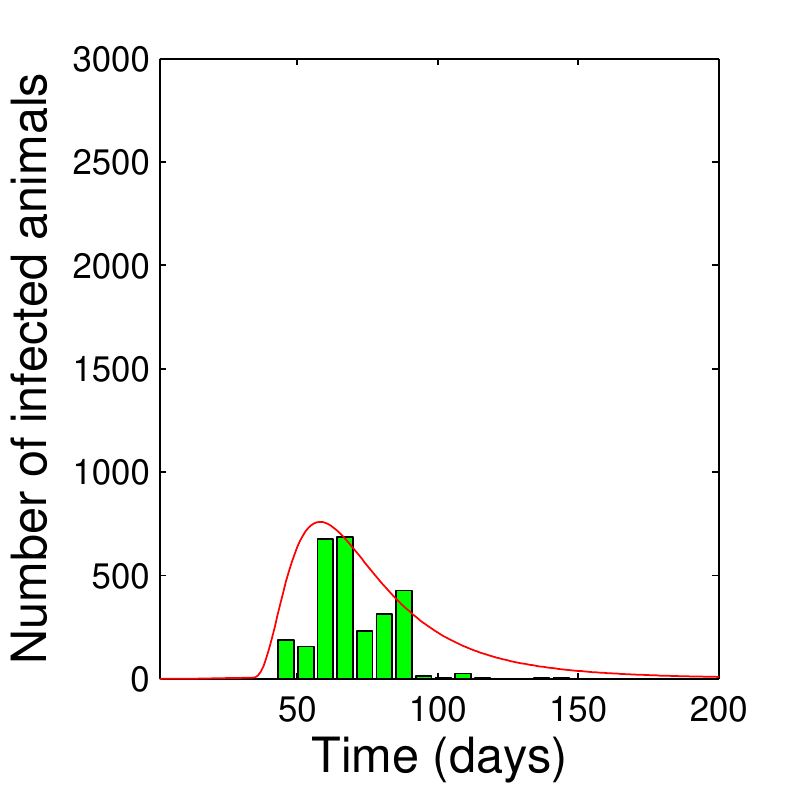}}
\hspace{0.0001in}
\subfigure[Simulation result and incidence data for humans in Northern Cape Province]{
\includegraphics[angle=0,width=6.5cm,height=5.2cm]{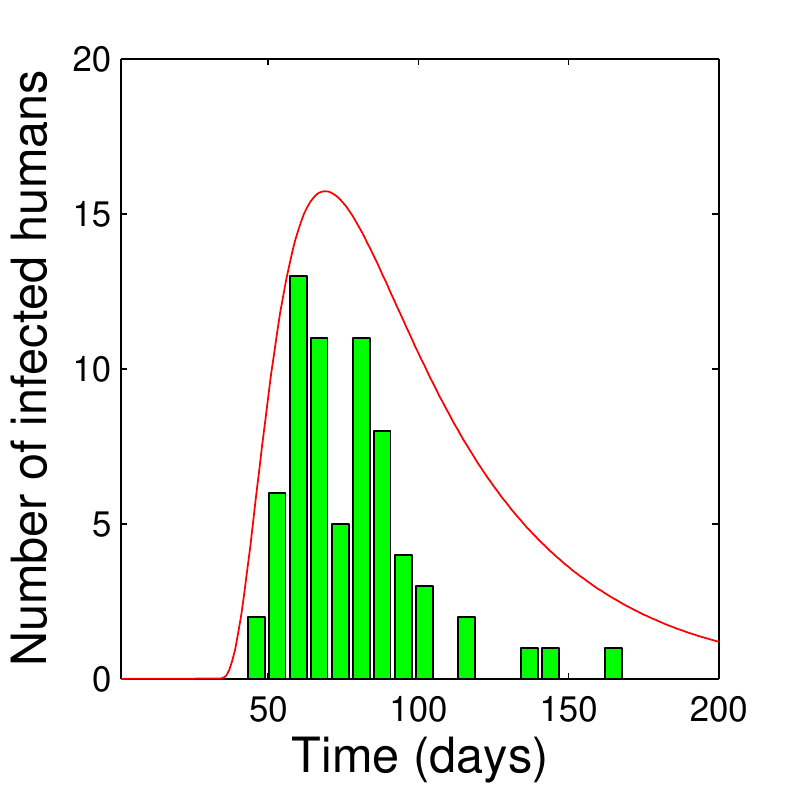}}
\subfigure[Simulation result and incidence data for sheep in Eastern Cape Province]{
\includegraphics[angle=0,width=6.5cm,height=5.2cm]{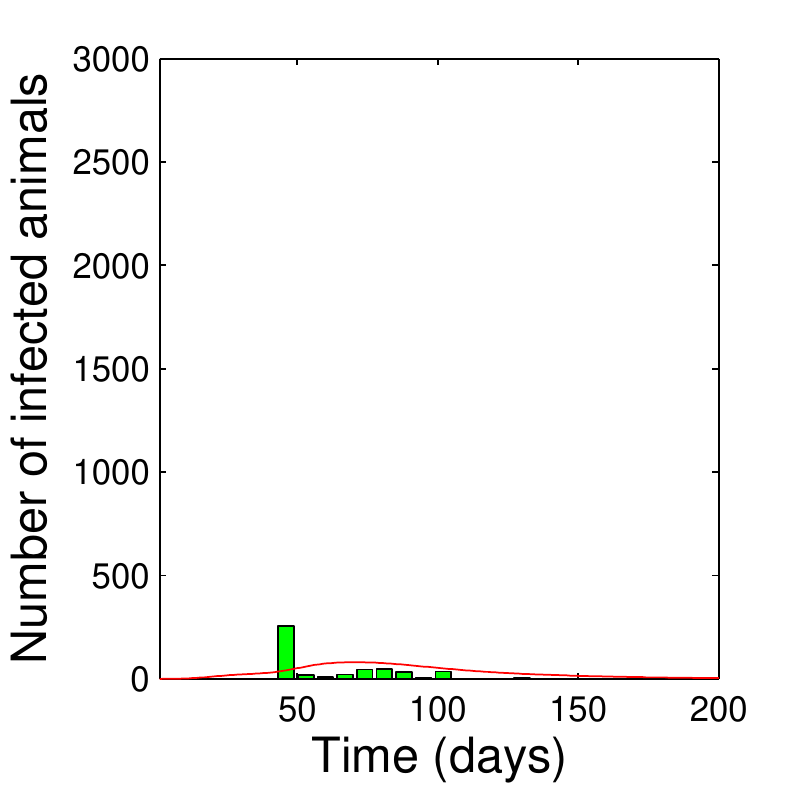}}
\hspace{0.0001in}
\subfigure[Simulation result and incidence data for humans in Eastern Cape Province]{
\includegraphics[angle=0,width=6.5cm,height=5.2cm]{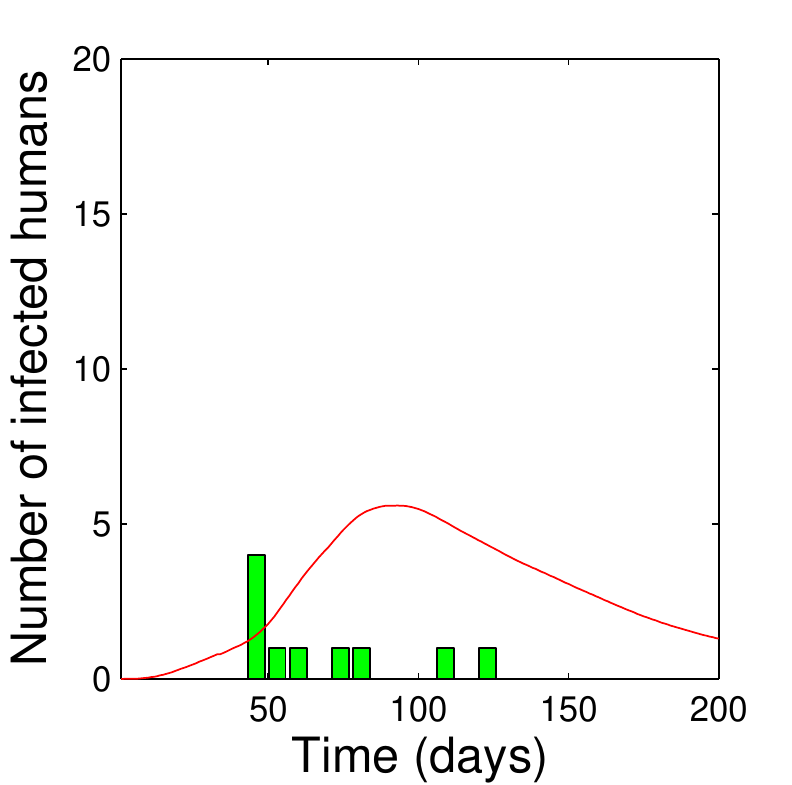}}
\caption{Simulation results and incidence data from January, $2010$ in South Africa (bars are data and lines are simulation results)}
\label{fig:casestudy}
\end{figure}

Outbreak data for animals are obtained from \cite{DiseaseBioPortal, OIE2010}, while outbreak data for human subpopulations are  collected from \cite{DepartmentAgriculturelivestock2009, NICD2010}. As far as animal data is concerned, we chose to analyze RVF incidence in the sheep population. Because the granularity of human incidence data is provided at  Province level, each node in the network represents a province. We selected three provinces: Free State (location $1$), Northern Cape (location $2$) and Eastern Cape (location $3$), because they had the highest levels of RVF incidence for humans. The curves of the incidence data are shown in Figure \ref{fig:casestudy} using green histograms, while the red curves represent simulations obtained with our model. From the data in Figure \ref{fig:casestudy}, it is possible to observe that the epidemic started first in the Free State Province and later in Northern Cape Province. The sustained heavy rainfall likely triggered the outbreak, causing infected eggs to hatch  in the Free State Province. Additionally,  the number of animal and human cases in Eastern Cape Province is smaller than the other two provinces.

\subsection{Sensitivity Analysis}\label{section:sensitivityandstability}
The three parameters $c_1$, $c_2$ and $c_3$ are estimated using the least square approach, which is based on minimization of errors between the incidence data of  humans and the percentage of humans calculated by the mathematical model. At first, we establish an objective function.  At each sample time, we calculate  the difference between the number of humans calculated by differential equations and that reported  \cite{Sahealth2010} during outbreaks in three provinces of South Africa from January, $2010$. We calculate the square of each difference. Then, we add all the squares for each location in each day together to obtain the objective function as is shown below.  Minimization of the objective function is initiated by providing initial values $c_{10}$, $c_{20}$ and $c_{30}$ for each parameter. The differential equations are solved with each set of the parameters and the square errors between the number of infected humans obtained from the objective function and those from incidence data are calculated. The  parameters $c_1$, $c_2$ and $c_3$ we used in the model are $c_1=0.009$, $c_2=0.05$ and $c_3=0.005$.

\begin{align}
F=\sum^{t_f}_{t=t_0}\sum^n_{i=1}[(I_{4i}(t)-PR_{4i}(t))^2]
\end{align}
In the equations above,\\
$n=$ the number of nodes\\
$t_0=$starting time\\
$t_f=$end time\\
$I_{4i}(t)=$human prevalence calculated by the model\\
$PR_{4i}(t)=$human prevalence reported  \\

\begin{figure}[http!]
\centering
\subfigure[Relative error of the number of infected humans with different value of $c_1$]{
\includegraphics[angle=0,width=4.1cm,height=4.2cm]{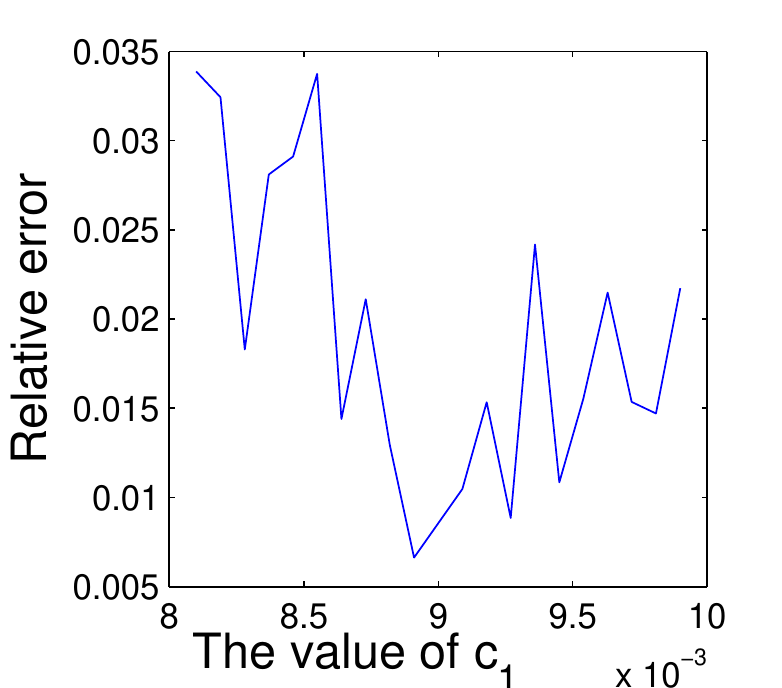}}
\hspace{0.0001in}
\subfigure[Relative error of the number of infected humans with different value of $c_2$]{
\includegraphics[angle=0,width=4.1cm,height=4.2cm]{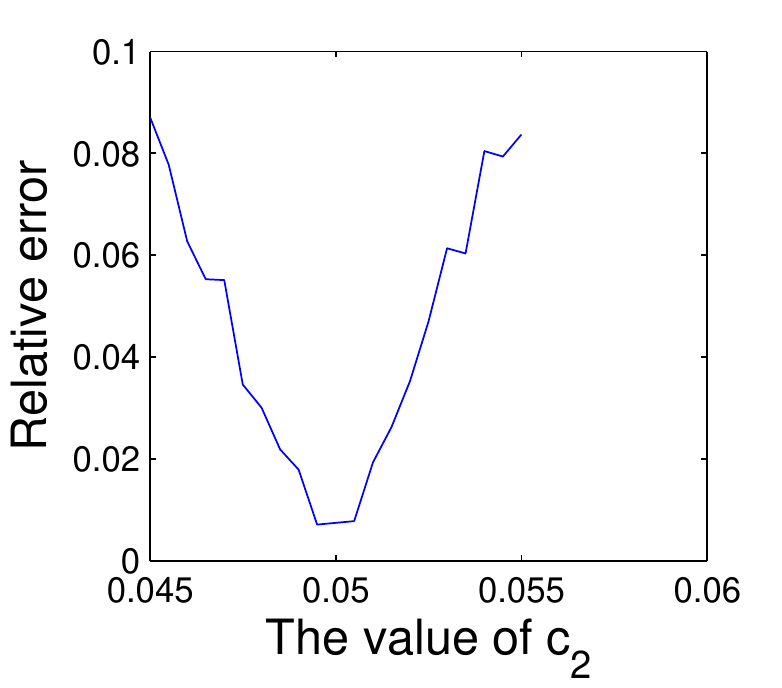}}
\hspace{0.0001in}
\subfigure[Relative error of the number of infected humans with different value of $c_3$]{
\includegraphics[angle=0,width=4.1cm,height=4.2cm]{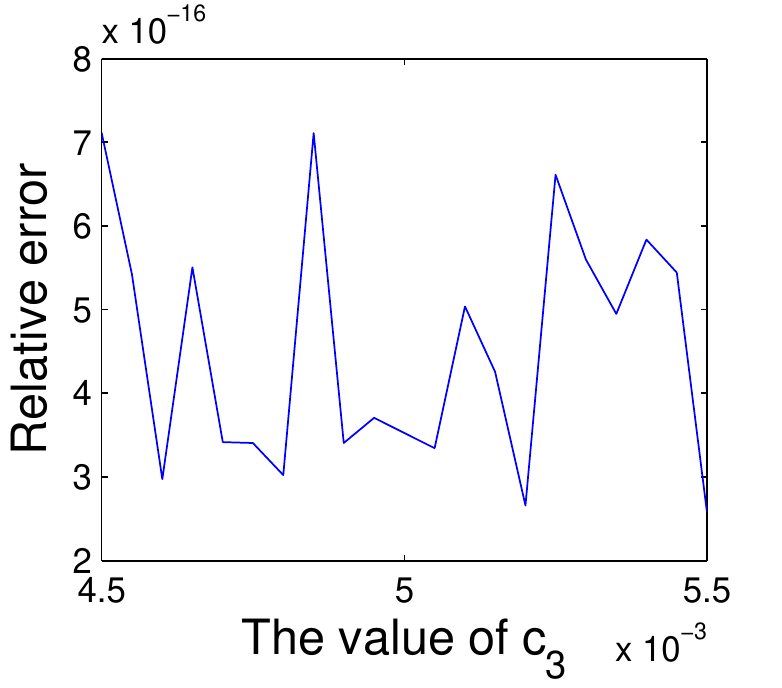}}
\caption{The relative error of the number of infected humans with changing one of the parameters $c_1$, $c_2$, and $c_3$}
\label{fig:relativeerrors}
\end{figure}

To conduct a sensitivity analysis of the parameters $c_1$, $c_2$, and $c_3$ in equations $(\ref{equation:mosquitoweight})$, (\ref{equation:animaloweight}), and $(\ref{equationhumanweight})$, we have changed each parameter within $\pm 10\%$ of the values $c_1=0.009$, $c_2=0.05$ and $c_3=0.005$, keeping the other parameters constant. This analysis allows an evaluation of the impact of uncertainties in the parameter estimations. The percentage of infected humans obtained from simulation with $c_1=0.009$, $c_2=0.05$ and $c_3=0.005$  is represented as $I_{Oi}(t)$. The percentage of infected human obtained from simulation with the  parameters within $\pm 10\%$ bound  is represented as $I_{4i}(t)$. The relative errors between the fractions of infected humans are calculated for each set of parameters, in each location, at time $t$ as $|\frac{I_{4i}(t)-I_{Oi}(t)}{I_{4i}(t)}|$.
The relative errors and the lower bound and upper bound of the parameters are shown in Figure \ref{fig:relativeerrors}.

All the values of relative errors shown in  Figure \ref{fig:relativeerrors} are smaller than $10\%$, proving the model robustness with respect to limited uncertainties in the parameter estimation. The rest of the parameters such as contact rate $\beta_{12}$, $\beta_{21}$, $\beta_{23}$, $\beta_{32}$, death rate $d_1$, $d_3$ and recovery rate $\gamma_2$ are  the most significant parameters in \cite{Gaff2007}. Similarly, $\beta_{14}$, $\beta_{24}$, $\beta_{34}$ and $\gamma_4$ are also the most significant parameters in this model.

\subsection{Analysis of Simulation Results}\label{section:simulationresults}
\begin{figure}[h]
\centering
\subfigure[Free State province]{
\includegraphics[angle=0,width=4.1cm,height=4.5cm]{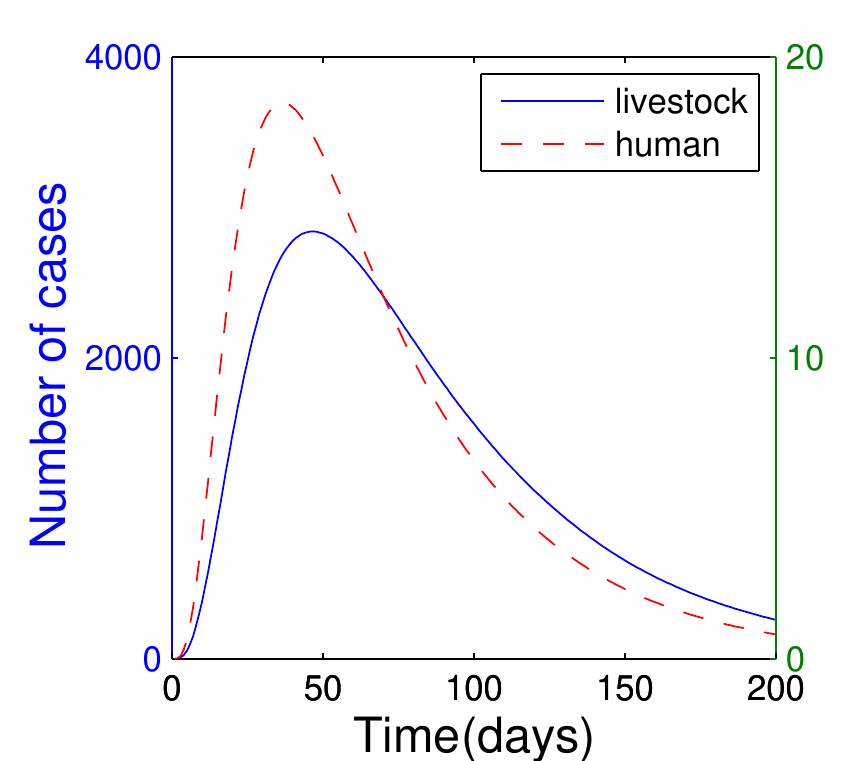}}
\hspace{0.1in}
\subfigure[Northern Cape province]{
\includegraphics[angle=0,width=4.1cm,height=4.5cm]{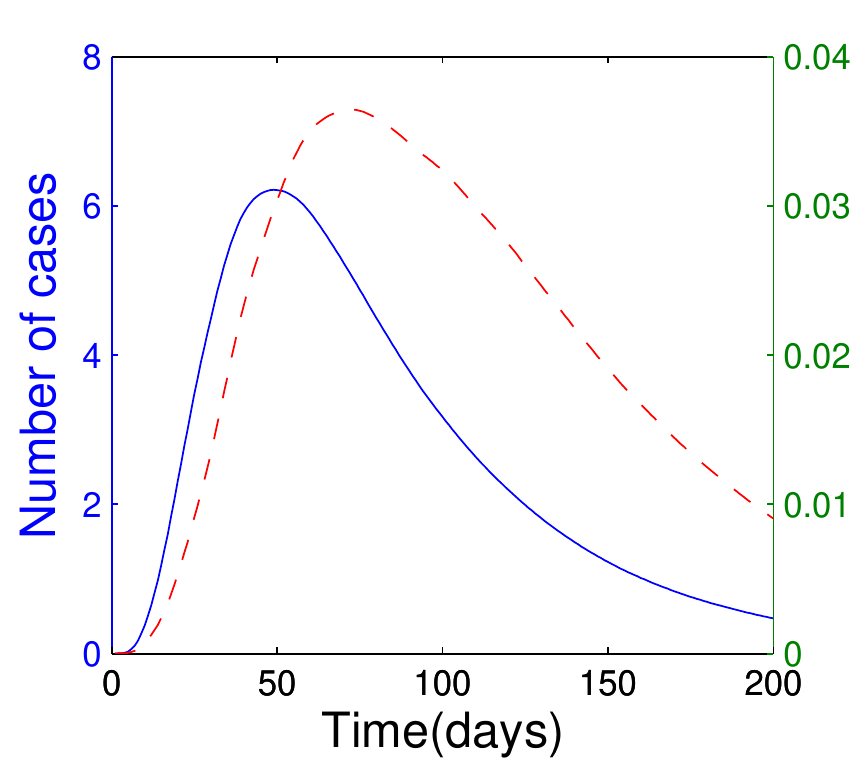}}
\hspace{0.1in}
\subfigure[Eastern Cape province]{
\includegraphics[angle=0,width=4.1cm,height=4.5cm]{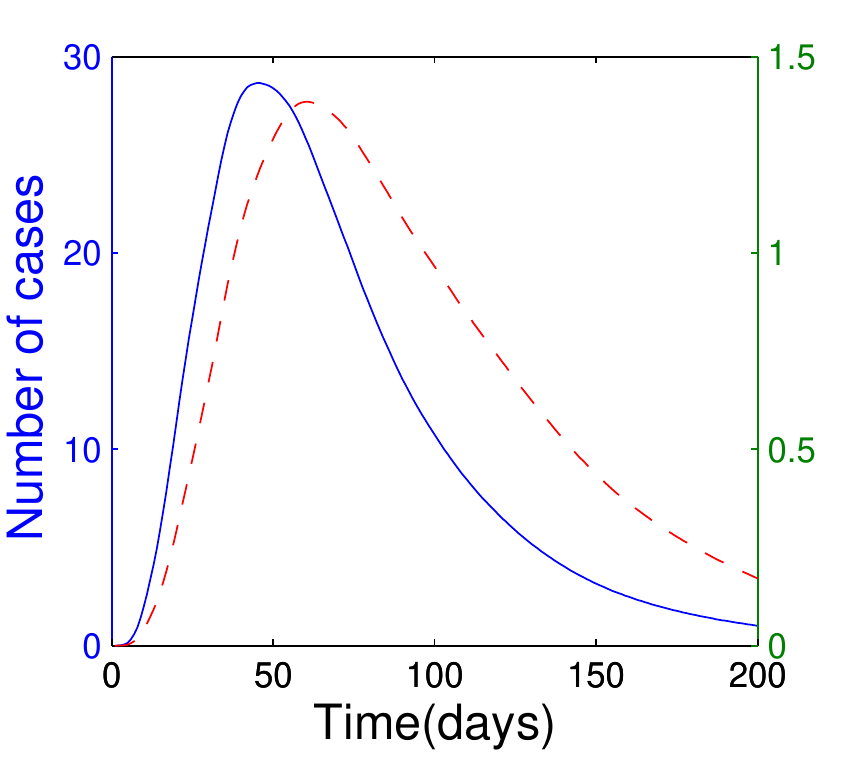}}
\caption{Simulation results with nonzero movement weights (the solid line represents livestock with y-axis on the left and the dash line represents humans with y-axis on the right)}
\label{fig:movement}
\end{figure}

To explore the behavior of RVFV, we conducted numerical simulations of an open system considering movement of the four species among different locations. To test the validity of the model, we changed some parameters in the weights to see the impact of each variation.
If the number of infected eggs $Q_{11}=10$, $Q_{12}=0$, and $Q_{13}=0$, at the beginning infected eggs only exist in location $1$, Free State. However, our model considers movement of mosquitoes to other locations with wind. As a consequence, infected animals and humans appear in all three locations as is shown in Figure \ref{fig:movement}. Therefore, the infection spreads due to movement of the four populations.

\begin{figure}[h]
\centering
\subfigure[Free State province]{
\includegraphics[angle=0,width=4.1cm,height=4.5cm]{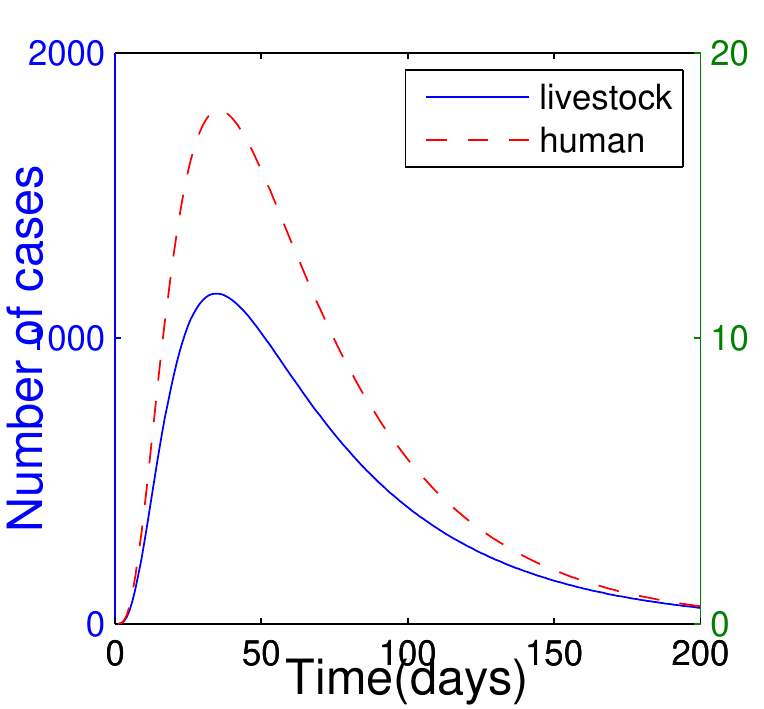}}
\hspace{0.1in}
\subfigure[Northern Cape province]{
\includegraphics[angle=0,width=4.1cm,height=4.5cm]{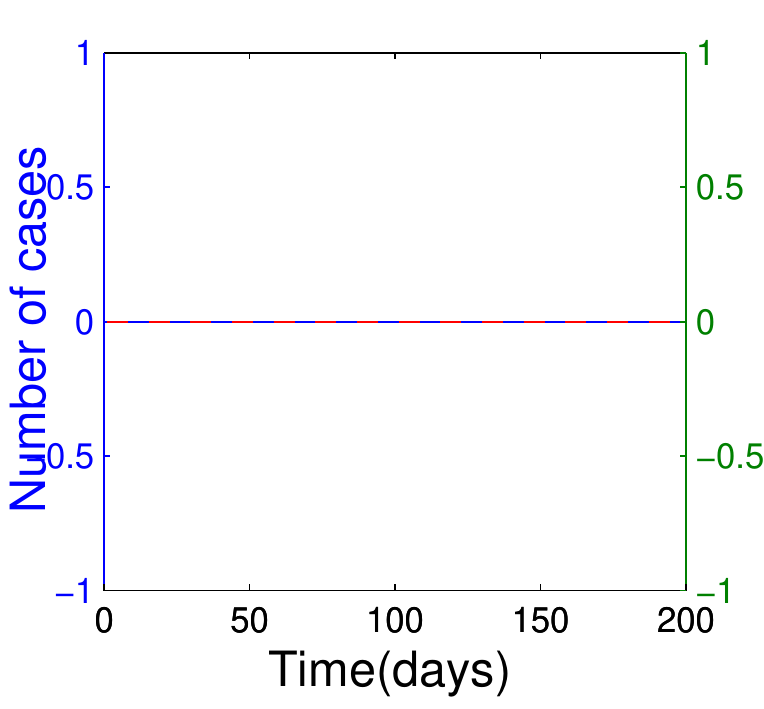}}
\hspace{0.1in}
\subfigure[Eastern Cape province]{
\includegraphics[angle=0,width=4.1cm,height=4.5cm]{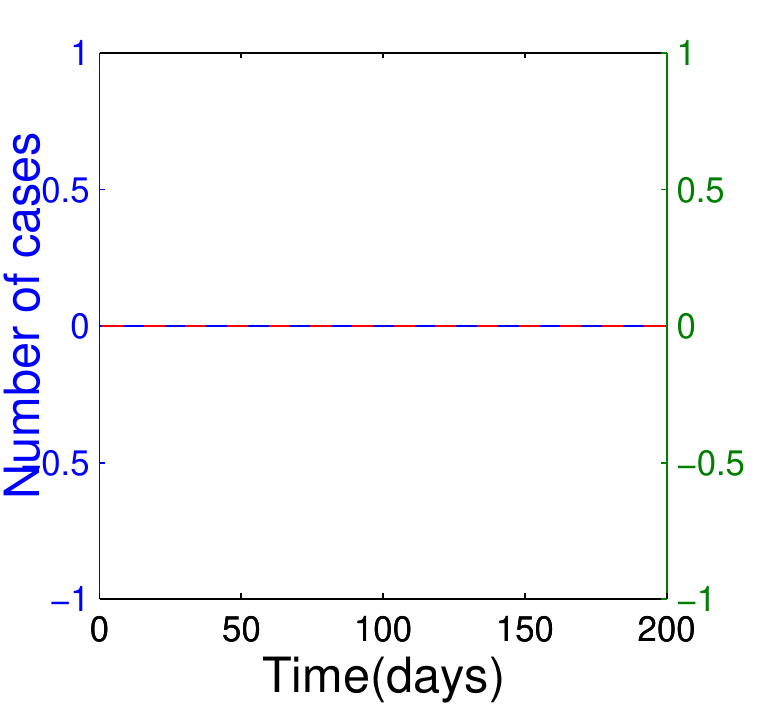}}
\caption{Simulation results with $c1=c2=c3=0$ (the solid line represents livestock with y-axis on the left and the dash line represents humans with y-axis on the right)}
\label{fig:nomovement}
\end{figure}

If we also assume that at the beginning infected eggs only exist in location $1$, $Q_{11}=10$, $Q_{12}=0$ and $Q_{13}=0$, and movements of the four species from one location to another are not allowed, $c_1=0$, $c_2=0$, and  $c_3=0$, then infected animals and humans will not appear in location $2$ and location $3$ as is shown in Figure \ref{fig:nomovement}. We can test the mitigation strategy of movement ban with this model.
We performed the simulations to reproduce the RVF outbreak in the three South African Provinces. The simulation results and the incidence data are shown in Figure \ref{fig:casestudy}. The model can differentiate the maximum number of infected individuals among the three different provinces, it also reproduces the different starting time of the outbreak in the three locations. With a homogeneous population model, such as the one in  \cite{Gaff2007}, the spatial differentiation is not possible.

The animal incidence curves provided by the model were always an overestimation of the data, since underreporting is very common during outbreaks.  Finally, our approach in which the fractions of each subpopulation in each compartment are expressed as continuous variables,  requires a large number of cases to be accurate. For this reason, the incidence data for location $3$, Eastern Cape, are better approximated by a stochastic model. The model has shown the ability  of fitting the data. The starting time and trend of outbreak dynamics have been reproduced by the model.

\section{Conclusions}\label{section:conclusionanddiscussion}
A meta-population, network-based, deterministic RVF model is presented here. The animal, human and mosquito movement and their spatial distribution are considered by the model. The model successfully describes a real outbreak dynamics of RVFV, taking into account space and movement. When considering $n$ locations or nodes $(n >= 1)$, there are $21*n$ differential equations and $21*n$ variables in our model, while there are only $14$ equations with $14$ variables in the model presented in \cite{ Mpeshe2011} and $16$ equations  with $16$ variables presented in \cite{Gaff2007}. Greater accuracy of our model is obtained at the cost of an increased complexity. The novelty of our model is that it considers a weighted contact network to represent the movement of four species. Subpopulations at the node level are also incorporated in our model. Additionally, parameters representing mosquito propagation and development are not constant but are the functions of climate factors. The model has been evaluated using data from the recent outbreak in South Africa. We reproduced not only the starting time but also  the trend of RVFV transmission with time in different locations. The model  has shown to be very promising notwithstanding the limitation of the data.  Due to the flexibility and accuracy of the  proposed model, we can test and design multiple and different mitigation strategies in different locations at different times.
The lower bound and upper bound of the reproduction number for homogeneous populations are shown to be very close to the exact value, and they provide insights on the biology of the spreading process.
Future work in follow-up mathematical models includes the development of a stochastic model, the study of the impact of climate changes on the epidemiology and control of RVF, and the improvement of the mosquito movement model considering diffusion equations. Moreover, the  carrying capacities of mosquitoes will be considered dependent on climate factors in the future.

\section*{Acknowledgments}
This work has been supported by the DHS Center of Excellence for Emerging and Zoonotic Animal Diseases (CEEZAD), and by the National Agricultural Biosecurity Center (NABC) at Kansas State University. We would like to give special thanks to the anonymous editor and reviewer for their  comments. We are grateful to Jason Coleman, Regina M. Beard, Livia Olsen, and Kelebogile Olifant for their help on bibliography research. We would  like to give thanks to Duygu Balcan, Phillip Schumm, Faryad Darabi Sahneh, Anton Lyubinin, and Getahun Agga for the help in producing the paper, and Kristine Bennett for answering questions on entomology.
\section*{Appendix}

\subsection* {Exact Computation of  $R_0$}

We compute $R_0$ as the spectral radius of the next generation matrix of the entire system \cite{diekmann2000mathematical, Mpeshe2011}, $R_0=\rho (FV^{-1})$.  Before applying the method we need to verify that the five assumptions in \cite{van2002reproduction} are satisfied \cite{Kim2010}. First, the equations in the system are reordered so that the first $m \ (m=9)$ compartments correspond to infected individuals.

\allowdisplaybreaks
\begin{align}
  \frac{\dif Q_{1}}{\dif t} &= b_1 q_1I_{1} -\theta_1Q_{1} \\
  \frac{\dif E_{1}}{\dif t} &=\beta_{21}S_{1}I_{2}/N_{2}-\varepsilon_1E_{1}-d_1E_{1}N_{1}/K_1\\
 \frac{\dif I_{1}}{\dif t} &=\theta_1Q_{1}+\varepsilon_1E_{1}-d_1I_{1}N_{1}/K_1\\
  \frac{\dif E_{2}}{\dif t} &=\beta_{12}S_{2}I_{1}/N_{1}
+\beta_{32}S_{2}I_{3}/N_{3}-\varepsilon_2E_{2}-d_2E_{2}N_{2}/K_{2}\\
  \frac{\dif I_{2}}{\dif t} &=\varepsilon_2E_{2}-\gamma_2I_{2}-\mu_2I_{2}-d_2I_{2}N_{2}/K_{2}\\
  \frac{\dif E_{3}}{\dif t} &=\beta_{23}S_{3}I_{2}/N_{2}-\varepsilon_3E_{3}-d_3E_{3}N_{3}/K_3\\
  \frac{\dif I_{3}}{\dif t} &=\varepsilon_3E_{3}-d_3I_{3}N_{3}/K_3\\
 \frac{\dif E_{4}}{\dif t} &=\beta_{14}S_{4}I_{1}/N_{1}+f\beta_{24}S_{4}I_{2}/N_{2}+
\beta_{34}S_{4}I_{3}/N_{3}-d_4E_{4}N_{4}/K_{4}-\varepsilon_4E_{4}\\
  \frac{\dif I_{4}}{\dif t} &=\varepsilon_4E_{4}-\gamma_4I_{4}-\mu_4I_{4}-d_4I_{4}N_{4}/K_{4}\\
  \frac{\dif P_{1}}{\dif t} &= b_1 \left(N_{1}-q_1I_{1} \right) -\theta_1P_{1}\\
 \frac{\dif P_{3}}{\dif t} &=b_3N_{3}-\theta_3P_{3}\\
 \frac{\dif S_{1}}{\dif t} &=\theta_1P_{1}-\beta_{21}S_{1}I_{2}/N_{2}-d_1S_{1}N_{1}/K_1\\
\frac{\dif S_{2}}{\dif t} &=b_2N_{2}-d_2S_{2}N_{2}/K_{2}-\beta_{12}S_{2}I_{1}/N_{1}
-\beta_{32}S_{2}I_{3}/N_{3} \\
 \frac{\dif S_{3}}{\dif t} &=\theta_3P_{3}-\beta_{23}S_{3}I_{2}/N_{2}-d_3S_{3}N_{3}/K_3\\
  \frac{\dif S_{4}}{\dif t} &=b_4N_{4}
-\beta_{14}S_{4}I_{1}/N_{1}-f\beta_{24}S_{4}I_{2}/N_{2}-
\beta_{34}S_{4}I_{3}/N_{3}-d_4S_{4}N_{4}/K_{4}\\
\frac{\dif R_{2}}{\dif t} &=\gamma_2I_{2}-d_2R_{2}N_{2}/K_{2}\\
 \frac{\dif R_{4}}{\dif t} &=\gamma_4I_{4}-d_4R_{4}N_{4}/K_{4}
\end{align}
\newline
The above system can be written as $f_k(x)=\mathscr{F}_k(x)- \mathscr{V}_k(x)$, $k=17$\\

where $$\begin{array} {lllllllllllllllll}
x=[Q_1 &  E_1 & I_1 & E_2& I_2  & E_3 & I_3 &
E_4 &  I_4 & P_1 &  P_3  & S_1 & S_2 & S_3 & S_4  & R_2 & R_4]^T,%
\end{array}$$  is the number of individuals in each compartment.\\
and $$\begin{array} {rlllllllllllllllll}X_S=&[Q^0_1 &  E^0_1 & I^0_1 & E^0_2 & I^0_2 &  E^0_3 &  I^0_3 & E^0_4 &  I^0_4 & P^0_1 & P^0_3 &  S^0_1 & S^0_2 & S^0_3 &  S^0_4 &  R^0_2 &R^0_4 ]^T\\
=&[0& 0 & 0& 0 & 0 & 0 & 0 & 0&  0& \frac{b_1^2K_1}{d_1 \theta_1}& \frac{b_3^2K_3}{d_3 \theta_3}& \frac{b_1K_1}{d_1}& \frac{b_2K_2}{d_2}&  \frac{b_3K_3}{d_3}& \frac{b_4K_4}{d_4}& 0& 0 ]
^T,%
\end{array}$$  is the set of  disease free states.\\
$\mathscr{F}(x)$, $ \mathscr{V}^-(x)$, and $\mathscr{V}^+(x)$ are given in the following.\\
$$\begin{array} {lllll}
 \mathscr{F}_1=&b_1q_1I_1, \ &\mathscr{V}_1^-=&\theta_1Q_{1}&\\
 \mathscr{F}_2=&\beta_{21}S_{1}I_{2}/N_{2}, \  &\mathscr{V}_2^-=&\varepsilon_1E_{1}+d_1E_{1}N_{1}/K_1&\\
\mathscr{F}_3=&0, \ & \mathscr{V}_3^-=&d_1I_{1}N_{1}/K_1, \  &\mathscr{V}_3^+=\theta_1Q_{1}+\varepsilon_1E_{1}\\
 \mathscr{F}_4=&\beta_{12}S_{2}I_{1}/N_{1}+\beta_{32}S_{2}I_{3}/N_{3}, \ &\mathscr{V}_4^-=&\varepsilon_2E_{2}+d_2E_{2}N_{2}/K_{2}&\\
 \mathscr{F}_5=&0, & \mathscr{V}_5^-=&\gamma_2I_{2}+\mu_2I_{2}+d_2I_{2}N_{2}/K_{2}, \  &\mathscr{V}_5^+=\varepsilon_2E_{2} \  \\
\mathscr{F}_6=&\beta_{23}S_{3}I_{2}/N_{2},  \ &\mathscr{V}_6^-=&\varepsilon_3E_{3}+d_3E_{3}N_{3}/K_3&\\
 \mathscr{F}_7=&0,  \  &\mathscr{V}_7^-=&d_3I_{3}N_{3}/K_3\ &\mathscr{V}_7^+=\varepsilon_3E_{3},\\
 \mathscr{F}_8=&\beta_{14}S_{4}I_{1}/N_{1}+f\beta_{24}S_{4}I_{2}/N_{2}, \  &\mathscr{V}_8^-=&d_4E_{4}N_{4}/K_{4}+\varepsilon_4E_{4}&\\
&+\beta_{34}S_{4}I_{3}/N_{3}, &&\\
 \mathscr{F}_9=&0,  \  &\mathscr{V}_9^-=&\gamma_4I_{4}+\mu_4I_{4}+d_4I_{4}N_{4}/K_{4},  \ &\mathscr{V}_9^+=\varepsilon_4E_{4}\\
\mathscr{F}_{10}=&0 , \ &\mathscr{V}_{10}^-=&b_1q_1I_1+\theta_1P_{1},  \  &\mathscr{V}_{10}^+= b_1 N_{1}\\
 \mathscr{F}_{11}=&0,\  &\mathscr{V}_{11}^-=&\theta_3P_{3}, \  &\mathscr{V}_{11}^+=b_3N_{3}\\
 \mathscr{F}_{12}=&0, \ &\mathscr{V}_{12}^-=&\beta_{21}S_{1}I_{2}/N_{2}+d_1S_{1}N_{1}/K_1, \  &\mathscr{V}_{12}^+=\theta_1P_{1}\\
\mathscr{F}_{13}=&0, \  &\mathscr{V}_{13}^-=&d_2S_{2}N_{2}/K_{2}+\beta_{12}S_{2}I_{1}/N_{1}
+\beta_{32}S_{2}I_{3}/N_{3}, \  &\mathscr{V}_{13}^+=b_2N_{2}\\
\mathscr{F}_{14}=&0, \  &\mathscr{V}_{14}^-=&\beta_{23}S_{3}I_{2}/N_{2}+d_3S_{3}N_{3}/K_3, \  &\mathscr{V}_{14}^+=\theta_3P_{3}\\
 \mathscr{F}_{15}&=0, \ &\
\mathscr{V}_{15}^-=&=\beta_{14}S_{4}I_{1}/N_{1}+f\beta_{24}S_{4}I_{2}/N_{2},&\mathscr{V}_{15}^+=b_4N_{4}\\
& & &+
\beta_{34}S_{4}I_{3}/N_{3}+d_4S_{4}N_{4}/K_{4} \\
\mathscr{F}_{16}=&0, \  &\mathscr{V}_{16}^-=&d_2R_{2}N_{2}/K_{2}
 &\mathscr{V}_{16}^+=\gamma_2I_{2}\\
\mathscr{F}_{17}=&0, \ &\mathscr{V}_{17}^-=&d_4R_{4}N_{4}/K_{4},  \  &\mathscr{V}_{17}^+=\gamma_4I_{4}
\end{array}$$

As it can been easily seen, the following five assumptions \cite{van2002reproduction}  are satisfied.
\begin{list}{}{}
 \item (A1)  if $x\geqslant 0$, then $\mathscr{F}_i$, $\mathscr{V}_i^+$, $\mathscr{V}_i^-  \geqslant 0$ for $i= 1, ... ,17$.
 \item (A2)  if $x_i=0$, then $\mathscr{V}_i^-=0$; in particular, if $x \in X_s$, then  $\mathscr{V}_i^-=0$  for $i= 1,... ,9$.
 \item (A3)  $\mathscr{F}_i=0$ if $i>9$; there are no new infections in uninfected compartments.
 \item (A4)   if $x \in X_s$, then  $\mathscr{F}_i(x)=0 $ and $\mathscr{V}_i^+(x)=0$ for $i= 1, ... ,9$.
 \item (A5)   if  $\mathscr{F}(x)$ is set to $0$, then all eigenvalues of $Df(x_0)$ have negative real parts.
 \end{list}

To construct the next generation matrix, we only consider infected and
exposed compartments. The equations are transformed as follows.\newline

\begin{equation*}
\frac{d}{dt}%
\begin{bmatrix}
Q_{1} \\
E_{1} \\
I_{1} \\
E_{2} \\
I_{2} \\
E_{3} \\
I_{3} \\
E_{4} \\
I_{4}%
\end{bmatrix}
= \mathscr{F} - \mathscr{V} =
\begin{bmatrix}
b_1q_1I_{1} \\
\beta_{21}S_{1}I_{2}/N_{2} \\
0 \\
\beta_{12}S_{2}I_{1}/N_{1}+\beta_{32}S_{2}I_{3}/N_{3} \\
0 \\
\beta_{23}S_{3}I_{2}/N_{2} \\
0 \\
\beta_{14}S_{4}I_{1}/N_{1}+f\beta_{24}S_{4}I_{2}/N_{2}+%
\beta_{34}S_{4}I_{3}/N_{3} \\
0 \\
\end{bmatrix}
-
\begin{bmatrix}
\theta_1Q_{1} \\
d_1E_{1}N_1/K_1+\varepsilon_1E_{1} \\
-\theta_1Q_{1}+d_1I_{1}N_1/K_1-\varepsilon_1E_{1} \\
d_2E_{2}N_2/K_2+\varepsilon_2E_{2} \\
-\varepsilon_2E_{2} + d_2I_{2}N_2/K_2+\gamma_2I_{2}+\mu_2I_{2} \\
d_3E_{3}N_3/K_3+\varepsilon_3E_{3} \\
d_3I_{3}N_3/K_3-\varepsilon_3E_{3} \\
d_4E_{4}N_4/K_4+\varepsilon_4E_{4} \\
d_4I_{4}N_4/K_4-\varepsilon_4E_{4}+\gamma_4I_4+\mu_4I_4 \\
\end{bmatrix}%
,
\end{equation*}

\allowdisplaybreaks

The equation system is nonlinear; we linearize it,  deriving the two Jacobian matrices. First, the partial derivative of $\mathscr{F}$
with respect to each variable at the disease free equilibrium is as follows \cite%
{van2002reproduction}.\newline
\newline
\begin{equation}
F= \left[ {%
\begin{array}{cccccccccccc}
0 & 0 & b_1q_1 & 0 & 0 & 0 & 0 & 0 & 0  \\
0 & 0 & 0 & 0 & \beta_{21}\frac{S^0_{1}}{N^0_2} & 0 & 0 & 0 & 0   \\
0 & 0 & 0 & 0 & 0 & 0 & 0 & 0 & 0  \\
0 & 0 & \beta_{12}\frac{S^0_{2}}{N^0_1} & 0 & 0 & 0 & \beta_{32}\frac{S^0_{2}%
}{N^0_3} & 0 & 0  \\
0 & 0 & 0 & 0 & 0 & 0 & 0 & 0 & 0 \\
0 & 0 & 0 & 0 & \beta_{23}\frac{S^0_{3}}{N^0_2} & 0 & 0 & 0 & 0 \\
0 & 0 & 0 & 0 & 0 & 0 & 0 & 0 & 0  \\
0 & 0 & \beta_{14}\frac{S^0_{4}}{N^0_1} & 0 & f\beta_{24}\frac{S^0_{4}}{N^0_2}
& 0 & \beta_{34}\frac{S^0_4}{N^0_3} & 0 & 0  \\
0 & 0 & 0 & 0 & 0 & 0 & 0 & 0 & 0 \\
\end{array}
} \right] \label{equation:F}
\end{equation}

Second, the partial derivative of $\mathscr{V}$ with respect to each variable at
disease free equilibrium is as follows.\newline
\begin{equation}
V= \left[ {%
\begin{array}{cccccccccccc}
\theta_1 & 0 & 0 & 0 & 0 & 0 & 0 & 0 & 0  \\
0 & b_1+\varepsilon_1  & 0 & 0 & 0 & 0 & 0 & 0 & 0  \\
-\theta_1 & -\varepsilon_1 & b_1 & 0 & 0 & 0 & 0 & 0 & 0  \\
0 & 0 & 0 & b_2+\varepsilon_2  & 0 & 0 & 0 & 0 & 0  \\
0 & 0 & 0 & -\varepsilon_2 & b_2+\gamma_2+\mu_2 & 0 & 0 & 0 & 0 \\
0 & 0 & 0 & 0 & 0 & b_3+\varepsilon_3 & 0 & 0 & 0 \\
0 & 0 & 0 & 0 & 0 & -\varepsilon_3 & b_3  & 0 & 0  \\
0 & 0 & 0 & 0 & 0 & 0 & 0 & b_4+\varepsilon_4 & 0  \\
0 & 0 & 0 & 0 & 0 & 0 & 0 & -\varepsilon_4 & b_4+\gamma_4+\mu_4   \\
\end{array}
} \right] \label{equation:V}
\end{equation}

The inverse of matrix $V$ is computed as follows.\newline
\begin{equation*}
V^{-1}= \left[ {%
\begin{array}{cccccccccccc}
\frac{1}{\theta_1} & 0 & 0 & 0 & 0 & 0 & 0 & 0 & 0   \\
0 & \frac{1}{b_1+\varepsilon_1} & 0 & 0 & 0 & 0 & 0 & 0 & 0 \\
\frac{1}{b_1} & \frac{\varepsilon_1}{b_1(b_1+\varepsilon_1)} & \frac{1}{b_1} & 0 & 0 & 0 & 0 & 0&0 \\
0 & 0 & 0 & \frac{1}{b_2+\varepsilon_2} & 0 & 0 & 0 & 0 & 0 \\
0 & 0 & 0 & \frac{\varepsilon_2}{(b_2+\varepsilon_2)(b_2+\gamma_2+\mu_2)} & \frac{1}{b_2+\gamma_2+\mu_2} & 0 & 0 & 0 & 0
\\
0 & 0 & 0 & 0 & 0 & \frac{1}{b_3+\varepsilon_3} & 0 & 0 & 0   \\
0 & 0 & 0 & 0 & 0 & \frac{\varepsilon_3 }{b_3 (b_3+\varepsilon_3)} & \frac{1}{b_3} & 0 & 0
\\
0 & 0 & 0 & 0 & 0 & 0 & 0 & \frac{1}{b_4+\varepsilon_4} & 0   \\
0 & 0 & 0 & 0 & 0 & 0 & 0 & \frac{\varepsilon_4}{(b_4+\varepsilon_4)(b_4+\gamma_4+\mu_4)} & \frac{1}{b_4+\gamma_4+\mu_4}
\\
\end{array}
} \right]
\end{equation*}

Finally, the next generation matrix, which is the product $FV^{-1}$,  is as follows.%
\newline

\begin{equation*}
FV^{-1}= \left[ {%
\begin{array}{cccccccccccc}
A_1 & B_1 & A_1 & 0 & 0 & 0 & 0 & 0 & 0 \\
0 & 0 & 0 & C_1 & D_1 & 0 & 0 & 0 & 0  \\
0 & 0 & 0 & 0 & 0 & 0 & 0 & 0 & 0\\
E_1 & F_1 & G_1 & 0 & 0 & H_1 & I_1 & 0 & 0   \\
0 & 0 & 0 & 0 & 0 & 0 & 0 & 0 & 0 \\
0 & 0 & 0 & J_1 & L_1 & 0 & 0 & 0 & 0   \\
0 & 0 & 0 & 0 & 0 & 0 & 0 & 0 & 0   \\
M_1 & N_1 & P_1 & Q_1 & R_1 & S_1 & T_1 & 0 & 0  \\
0 & 0 & 0 & 0 & 0 & 0 & 0 & 0 & 0   \\
\end{array}
} \right]
\end{equation*}
where:\newline
\begin{align*}
A_1 &=q_1 \\
B_1&=\frac{q_1\varepsilon_1}{b_1+\varepsilon_1%
} \\
C_1&= \frac{%
\varepsilon_2\beta_{21}b_1K_1d_2}{(b_2+\varepsilon_2)(b_2+\gamma_2+\mu_2
)d_1b_2K_2} \\
D_1&= \frac{\beta_{21}b_1K_1d_2}{%
(b_2+\gamma_2+\mu_2 )d_1b_2K_2} \\
E_1&= \frac{\beta_{12}b_2K_2d_1}{b_1^2d_2K_1}
\\
F_1&=\frac{\varepsilon_1%
\beta_{12}b_2K_2d_1}{(b_1+\varepsilon_1)b_1^2d_2K_1} \\
G_1&= \frac{\beta_{12}b_2K_2d_1}{b_1^2d_2K_1}
\\
H_1& = \frac{%
\varepsilon_3\beta_{32}b_2K_2d_3}{(b_3+\varepsilon_3)b_3^2d_2K_3} \\
I_1&=\frac{\beta_{32}b_2K_2d_3}{b_3^2d_2K_3}
\\
J_1&=\frac{%
\varepsilon_2b_3K_3d_2\beta_{23}}{(b_2+\varepsilon_2)(b_2+\gamma_2+\mu_2
)d_3b_2K_2} \\
L_1&= \frac{b_3K_3d_2\beta_{23}}{%
(b_2+\gamma_2+\mu_2)d_3b_2K_2} \\
M_1&=\frac{b_4K_4d_1\beta_{14}}{b_1^2d_4K_1}
\\
N_1&= \frac{\varepsilon_1%
\beta_{14}b_4K_4d_1}{(b_1+\varepsilon_1)b_1^2d_4K_1} \\
P_1&=\frac{b_4d_1K_4\beta_{14}}{b_1^2d_4K_1}
\\
Q_1&=\frac{%
f\varepsilon_2b_4K_4d_2\beta_{24}}{(b_2+\varepsilon_2)(b_2+\gamma_2+%
\mu_2)d_4b_2K_2} \\
R_1&=\frac{fb_4K_4d_2\beta_{24}}{%
(b_2+\gamma_2+\mu_2)d_4b_2K_2} \\
S_1&=\frac{%
\varepsilon_3b_4K_4d_3\beta_{34}}{(b_3+\varepsilon_3 )b_3^2d_4K_3} \\
T_1&=\frac{b_4K_4d_3\beta_{34}}{b_3^2K_3d_4}
\end{align*}
Recall that the reproduction number is the spectral radius of $FV^{-1}$, we compute the nine eigenvalues $\lambda_i$ of $FV^{-1}$ to select the one with maximum magnitude.

\begin{equation*}
|FV^{-1}-\lambda I|= \left| {%
\begin{array}{cccccccccccc}
A_1-\lambda & B_1 & A_1 & 0 & 0 & 0 & 0 & 0 & 0  \\
0 & -\lambda & 0 & C_1 & D_1 & 0 & 0 & 0 & 0  \\
0 & 0 & -\lambda & 0 & 0 & 0 & 0 & 0 & 0 \\
E_1 & F_1 & G_1 & -\lambda & 0 & H_1 & I_1 & 0 & 0  \\
0 & 0 & 0 & 0 & -\lambda & 0 & 0 & 0 & 0\\
0 & 0 & 0 & J_1 & L_1 & -\lambda & 0 & 0 & 0 \\
0 & 0 & 0 & 0 & 0 & 0 & -\lambda & 0 & 0   \\
M_1 & N_1 & P_1 & Q_1 & R_1 & S_1 & T_1 & -\lambda & 0  \\
0 & 0 & 0 & 0 & 0 & 0 & 0 & 0 & -\lambda   \\
\end{array}
}\right|
\end{equation*}
\begin{align}
&=-\lambda^6[\lambda^3-A_1\lambda^2-(C_1F_1+J_1H_1)%
\lambda+(H_1A_1J_1+A_1C_1F_1-B_1C_1E_1)] \label{equation:P1}
\end{align}
\begin{align}
B_1C_1E_1&=\frac{q_1\varepsilon_1\varepsilon_2\beta_{12}\beta_{21}}{b_1(b_1+\varepsilon_1)(b_2+\varepsilon_2)(b_2+\mu_2+\gamma_2)} \label{equation:BCE}\\
A_1C_1F_1&=\frac{q_1\varepsilon_1\varepsilon_2\beta_{12}\beta_{21}}{b_1(b_1+\varepsilon_1)(b_2+\varepsilon_2)(b_2+\mu_2+\gamma_2)} \label{equation:ACF}
\end{align}

Because $B_1C_1E_1=A_1C_1F_1  $ as shown in Equation (\ref{equation:ACF}) and  (\ref{equation:BCE}) , equation (\ref{equation:P1}) can be rewritten as follows.\\
\begin{align}
-\lambda^6[\lambda^3-A_1\lambda^2-(C_1F_1+J_1H_1)%
\lambda+H_1A_1J_1] &=0  \label{equation:P2}
\end{align}

\begin{figure}[h]
\centering
\includegraphics[angle=0,width=6cm,height=5.7cm]{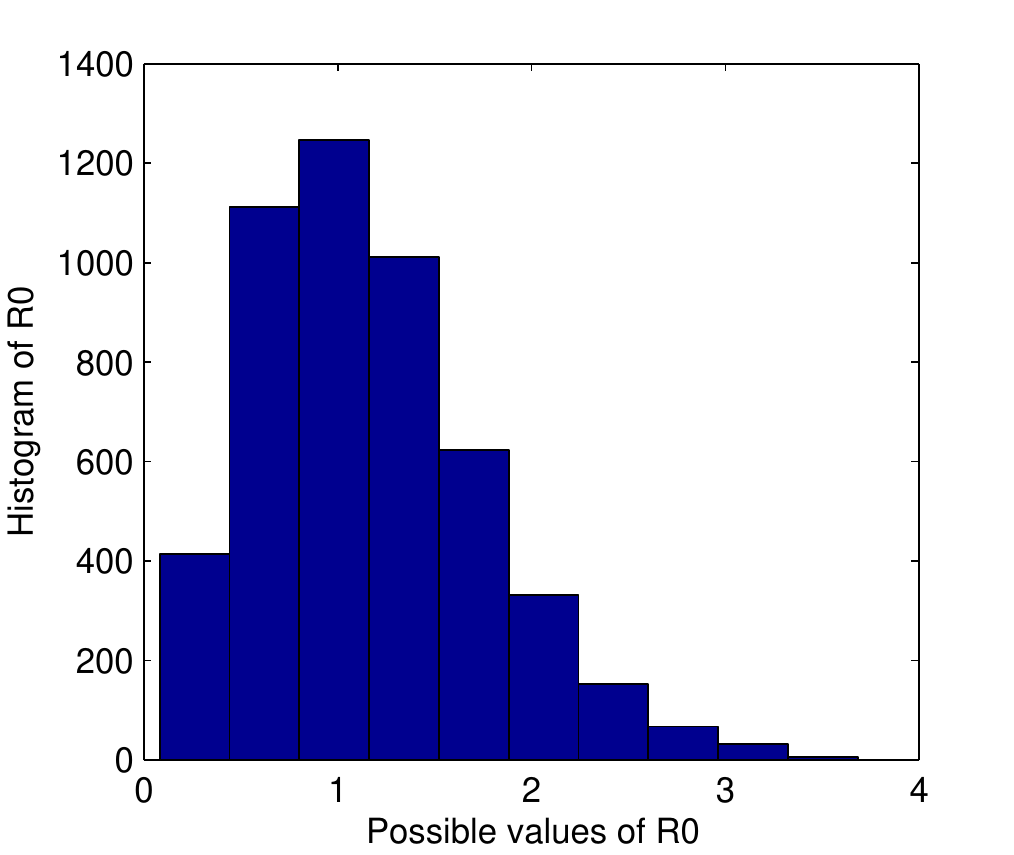}
\caption{Histogram of the reproduction number, the mean is $1.17$, the
maximum is $3.68$}
\label{fig:R0HV}
\end{figure}

Equation (\ref{equation:P2}) has six zero roots.  We only need to  solve the following equation to find $\max|\lambda_i| \ (i=1,2,3)$.\\
\begin{align}
\lambda^3-A_1\lambda^2-(C_1F_1+J_1H_1)
\lambda+H_1A_1J_1&=0  \label{equation:eigenvalue}
\end{align}
Equivalently,\\
\begin{align}
\lambda^3-q_1\lambda^2-[\frac{\varepsilon_1\varepsilon_2\beta_{21}\beta_{12}}{b_1(b_1+\varepsilon_1)(b_2+\varepsilon_2)(b_2+\mu_2+\gamma_2)}&+\frac{\varepsilon_2\varepsilon_3\beta_{23}\beta_{32}}{b_3(b_3+\varepsilon_3)(b_2+\varepsilon_2)(b_2+\mu_2+\gamma_2)}]\lambda\nonumber\\+\frac{q_1\varepsilon_2\varepsilon_3\beta_{23}\beta_{32}}{b_3(b_3+\varepsilon_3)(b_2+\varepsilon_2)(b_2+\mu_2+\gamma_2)}=0\label{equation:eigenvalue1}
\end{align}
We calculate the reproduction  number numerically with $5000$ different sets of parameters uniformly distributed within the range in \cite{Gaff2007}. The histogram of the reproduction number is shown in Figure \ref{fig:R0HV}. From the histogram, we can see that $R_0$ can be greater or smaller than $1$.
 In particular, the mean is $1.17$ and the maximum is $3.68$,
respectively. \\

\newpage

\subsection* {Upper and Lower Bound for $R_0$}
 Although we are able to only obtain the exact expression  of  $R_0$ numerically, we determine the lower bound and the upper bound  of $R_0$ in the following. \\
\begin{equation*}
|FV^{-1}-\lambda I|= \left| {%
\begin{array}{cccccccccccc}
A_1-\lambda & B_1 & A_1 & 0 & 0 & 0 & 0 & 0 & 0  \\
0 & -\lambda & 0 & C_1 & D_1 & 0 & 0 & 0 & 0  \\
0 & 0 & -\lambda & 0 & 0 & 0 & 0 & 0 & 0 \\
E_1 & F_1 & G_1 & -\lambda & 0 & H_1 & I_1 & 0 & 0  \\
0 & 0 & 0 & 0 & -\lambda & 0 & 0 & 0 & 0\\
0 & 0 & 0 & J_1 & L_1 & -\lambda & 0 & 0 & 0 \\
0 & 0 & 0 & 0 & 0 & 0 & -\lambda & 0 & 0   \\
M_1 & N_1 & P_1 & Q_1 & R_1 & S_1 & T_1 & -\lambda & 0  \\
0 & 0 & 0 & 0 & 0 & 0 & 0 & 0 & -\lambda   \\
\end{array}
}\right|=-\lambda^5 \left| {%
\begin{array}{cccccccccccc}
A_1-\lambda & B_1 & 0 & 0  \\
0 & -\lambda & C_1 & 0   \\
E_1 & F_1 & -\lambda & H_1  \\
0 & 0 & J_1 & -\lambda  \\
\end{array}
} \right|=-\lambda^5 | \mathcal{A}-\lambda I|
\end{equation*}

Where:
\begin{equation*}
 \mathcal{A}= \left[{%
\begin{array}{cccccccccccc}
A_1 &B_1 & 0 & 0  \\
0 &0 & C_1 & 0   \\
E_1 & F_1 &0 & H_1  \\
0 & 0 & J_1 &0\\
\end{array}
} \right]= \left[{%
\begin{array}{cccccccccccc}
q_1 &\frac{q_1\varepsilon_1}{b_1+\varepsilon_1%
} & 0 & 0  \\
0 &0 &\frac{%
\varepsilon_2\beta_{21}b_1K_1d_2}{(b_2+\varepsilon_2)(b_2+\gamma_2+\mu_2
)d_1b_2K_2}& 0   \\
 \frac{\beta_{12}b_2K_2d_1}{b_1^2d_2K_1}& \frac{\varepsilon_1%
\beta_{12}b_2K_2d_1}{(b_1+\varepsilon_1)b_1^2d_2K_1}&0 & \frac{%
\varepsilon_3\beta_{32}b_2K_2d_3}{(b_3+\varepsilon_3)b_3^2d_2K_3} \\
0 & 0 & \frac{%
\varepsilon_2b_3K_3d_2\beta_{23}}{(b_2+\varepsilon_2)(b_2+\gamma_2+\mu_2
)d_3b_2K_2} &0\\
\end{array}
} \right]
\end{equation*}
Matrix $FV^{-1} $ has five zero roots. To find out the spectral radius of $FV^{-1}$, we only need to find out the spectral radius of matrix $\mathcal{A}$ which can be rewritten as follows.\\
\begin{equation*}
 \mathcal{A}= \left[{%
\begin{array}{cccccccccccc}
0 &0& 0 & 0  \\
0 &0 &\frac{%
\varepsilon_2\beta_{21}b_1K_1d_2}{(b_2+\varepsilon_2)(b_2+\gamma_2+\mu_2
)d_1b_2K_2}& 0   \\
 \frac{\beta_{12}b_2K_2d_1}{b_1^2d_2K_1}& \frac{\varepsilon_1%
\beta_{12}b_2K_2d_1}{(b_1+\varepsilon_1)b_1^2d_2K_1}&0 & \frac{%
\varepsilon_3\beta_{32}b_2K_2d_3}{(b_3+\varepsilon_3)b_3^2d_2K_3} \\
0 & 0 & \frac{%
\varepsilon_2b_3K_3d_2\beta_{23}}{(b_2+\varepsilon_2)(b_2+\gamma_2+\mu_2
)d_3b_2K_2} &0\\
\end{array}
} \right]+ \left[{%
\begin{array}{cccccccccccc}
q_1 &\frac{q_1\varepsilon_1}{b_1+\varepsilon_1%
} & 0 & 0  \\
0 & 0 & 0& 0   \\
0 & 0 &0  & 0  \\
0 & 0 & 0 & 0 \\
\end{array}
} \right]= \mathcal{B}+ \mathcal{C}
\end{equation*}

$\mathcal{B}$  is the first matrix and $\mathcal{C}$ is the second matrix. The matrix $\mathcal{C}$  has three multiple eigenvalues $\lambda_2=\lambda_3=\lambda_4=0$. Since $Rank \ (\lambda_i I-\mathcal{C})|_{\lambda_2=\lambda_3=\lambda_4=0} =1$, there are  three linear independent eigenvectors  corresponding to  zero eigenvalue. Overall, matrix $\mathcal{C}$ has four linear independent eigenvectors. Therefore, matrix $\mathcal{C}$  can be diagonalized as follows.
\begin{align*}
\mathcal{C}&=  P\left[  {%
\begin{array}
[c]{cccccccc}%
q_1 &0&0&0\\
0 &0&0&0\\
0 &0&0&0\\
0 &0&0&0\\
\end{array}
}\right] P^{-1},
\end{align*}
where:\\
\[
P=%
\begin{bmatrix}
1 &-\frac{q_1\varepsilon_1}{b_1+\varepsilon_1}&0&0\\
0&q_1&0&0\\
0 &0&1&0\\
0 &0&0&1\\
\end{bmatrix},  \   \  \  \
P^{-1}=%
\begin{bmatrix}
1 &\frac{\varepsilon_1}{b_1+\varepsilon_1}&0&0\\
0&\frac{1}{q_1}&0&0\\
0 &0&1&0\\
0 &0&0&1\\
\end{bmatrix},
\]
Substitute $\mathcal{C}$ in the expression of $\mathcal{A}=\mathcal{B}+\mathcal{C}$.
\begin{align*}
\mathcal{A}&= P(P^{-1}\mathcal{B}P+ \begin{bmatrix}
q_1&0&0&0\\
0&0&0&0\\
0 &0&0&0\\
0 &0&0&0\\
\end{bmatrix})P^{-1}
= P(X+Y)P^{-1}
\end{align*}
where\\
\begin{equation*}
X= P^{-1}\mathcal{B}P=\begin{bmatrix}
0 &0&    \frac{\varepsilon_1\varepsilon_2\beta_{21}b_1K_1d_1}{(b_1+\varepsilon_1)(b_2+\varepsilon_2)(b_2+\gamma_2+\mu_2)d_1b_2K_2}&0\\
0&0&\frac{\varepsilon_2\beta_{21}b_1K_1d_2}{q_1(b_2+\varepsilon_2)(b_2+\gamma_2+\mu_2)d_1b_2K_2}&0\\
0 &\frac{\beta_{21}b_2K_2d_1}{b_1^2d_2K_1}&0&\frac{\varepsilon_3\beta_{32}b_2K_2d_3}{(b_3+\varepsilon_3)d_2b3^2K_3}\\
0 &0&\frac{\varepsilon_2\beta_{23}b_3K_3d_2}{(b_2+\varepsilon_2)(b_2+\gamma_2+\mu_2)d_3b_2K_2}&0\\
\end{bmatrix},Y=\begin{bmatrix}
q_1&0&0&0\\
0&0&0&0\\
0 &0&0&0\\
0 &0&0&0\\
\end{bmatrix}.
\end{equation*}

Matrix $Y$ is a $4 \times 4$ nonnegative diagonal matrix $(Y_{ij}=0 $ for $ i\neq j)$ and $0\leqslant Y_{ii} <q_1=\max_j v_{jj}<\infty ,( i=2,3,4)$, and matrix $X$ is nonnegative.
 Therefore, $\rho(X) \leqslant \rho(X+Y) \leqslant \rho(X)+q_1$ according to  Theorem $1.$  in \cite {Cohen1979}. The eigenvalues of matrix $X+Y$ are the same as those of $\mathcal A$ because the two matrices are similar. Similarly, the eigenvalues of
matrix $X$ are the same as those of matrix $\mathcal B$ .
\begin{align*}
\rho(X+Y)&=\rho(FV^{-1})=R_0\\
\rho(X)&= \sqrt{
\frac{\varepsilon_2}{(b_2+\varepsilon_2)
(b_2+\gamma_2+\mu_2)}
\Big[\frac{\varepsilon_1\beta_{12}\beta_{21}}{b_1
 (b_1+\varepsilon_1)}
+\frac{\varepsilon_3\beta_{32}\beta_{23}}{b_3(b_3+\varepsilon_3)}
\Big] }\\
\end{align*}

If we only count the horizontal transmission and denote the new $F$ (resp. $V)$ by $F_H$ (resp. $V_H)$ , $F_H$ and $V_H$ are as follows.\\
\begin{align*}
F_{H} &  =\left[  {%
\begin{array}
[c]{cccccccc}%
0 & 0 & 0 & \beta_{21}\frac{S_{1}^{0}}{N_{2}^{0}} & 0 & 0 & 0 & 0\\
0 & 0 & 0 & 0 & 0 & 0 & 0 & 0\\
0 & \beta_{12}\frac{S_{2}^{0}}{N_{1}^{0}} & 0 & 0 & 0 & \beta_{32}\frac
{S_{2}^{0}}{N_{3}^{0}} & 0 & 0\\
0 & 0 & 0 & 0 & 0 & 0 & 0 & 0\\
0 & 0 & 0 & \beta_{23}\frac{S_{3}^{0}}{N_{2}^{0}} & 0 & 0 & 0 & 0\\
0 & 0 & 0 & 0 & 0 & 0 & 0 & 0\\
0 & \beta_{14}\frac{S_{4}^{0}}{N_{1}^{0}} & 0 & \beta_{24}\frac{S_{4}^{0}%
}{N_{2}^{0}} & 0 & \beta_{34}\frac{S_{4}^{0}}{N_{3}^{0}} & 0 & 0\\
0 & 0 & 0 & 0 & 0 & 0 & 0 & 0
\end{array}
}\right]  ,\\
V_{H} &  =\left[
\begin{array}
[c]{cccccccc}%
\frac{d_{1}N_{1}^{0}}{K_{1}}+\varepsilon_{1} & 0 & 0 & 0 & 0 & 0 & 0 & 0\\
-\varepsilon_{1} & \frac{d_{1}N_{1}^{0}}{K_{1}} & 0 & 0 & 0 & 0 & 0 & 0\\
0 & 0 & \frac{d_{2}N_{2}^{0}}{K_{2}}+\varepsilon_{2} & 0 & 0 & 0 & 0 & 0\\
0 & 0 & -\varepsilon_{2} & \frac{d_{2}N_{2}^{0}}{K_{2}}+\gamma_{2}+\mu_{2} &
0 & 0 & 0 & 0\\
0 & 0 & 0 & 0 & \frac{d_{3}N_{3}^{0}}{K_{3}}+\varepsilon_{3} & 0 & 0 & 0\\
0 & 0 & 0 & 0 & -\varepsilon_{3} & \frac{d_{3}N_{3}^{0}}{K_{3}} & 0 & 0\\
0 & 0 & 0 & 0 & 0 & 0 & \frac{d_{4}N_{4}^{0}}{K_{4}}+\varepsilon_{4} & 0\\
0 & 0 & 0 & 0 & 0 & 0 & -\varepsilon_{4} & \frac{d_{4}N_{4}^{0}}{K_{4}}%
+\gamma_{4}+\mu_{4}%
\end{array}
\right]
\end{align*}
By calculation,
\begin{align*}
R_0^H=\rho(F_HV_H^{-1})=\sqrt{
\frac{\varepsilon_2}{(b_2+\varepsilon_2)
(b_2+\gamma_2+\mu_2)}
\Big[\frac{\varepsilon_1\beta_{12}\beta_{21}}{b_1
 (b_1+\varepsilon_1)}
+\frac{\varepsilon_3\beta_{32}\beta_{23}}{b_3(b_3+\varepsilon_3)}
\Big] }=\rho(X)\\
\end{align*}
Therefore,
\begin{align}
R_0^H \leqslant R_0\leqslant R_0^H+q_1 \label{equation:bound}
\end{align}

We denote $R_0^H$ and $R_0^H+q_1$ as $R_0^L$ and $R_0^U$, respectively. \\

\subsection*{Numerical Comparison among $R_0$,  $ R_0^L$, and  $ R_0^U$}

\begin{figure}[!htbp]
\centering
\subfigure[The reproduction number and its upper  bound.]{
\label{fig:upperbound}
\includegraphics[angle=0,width=10.9cm,height=10.9cm]{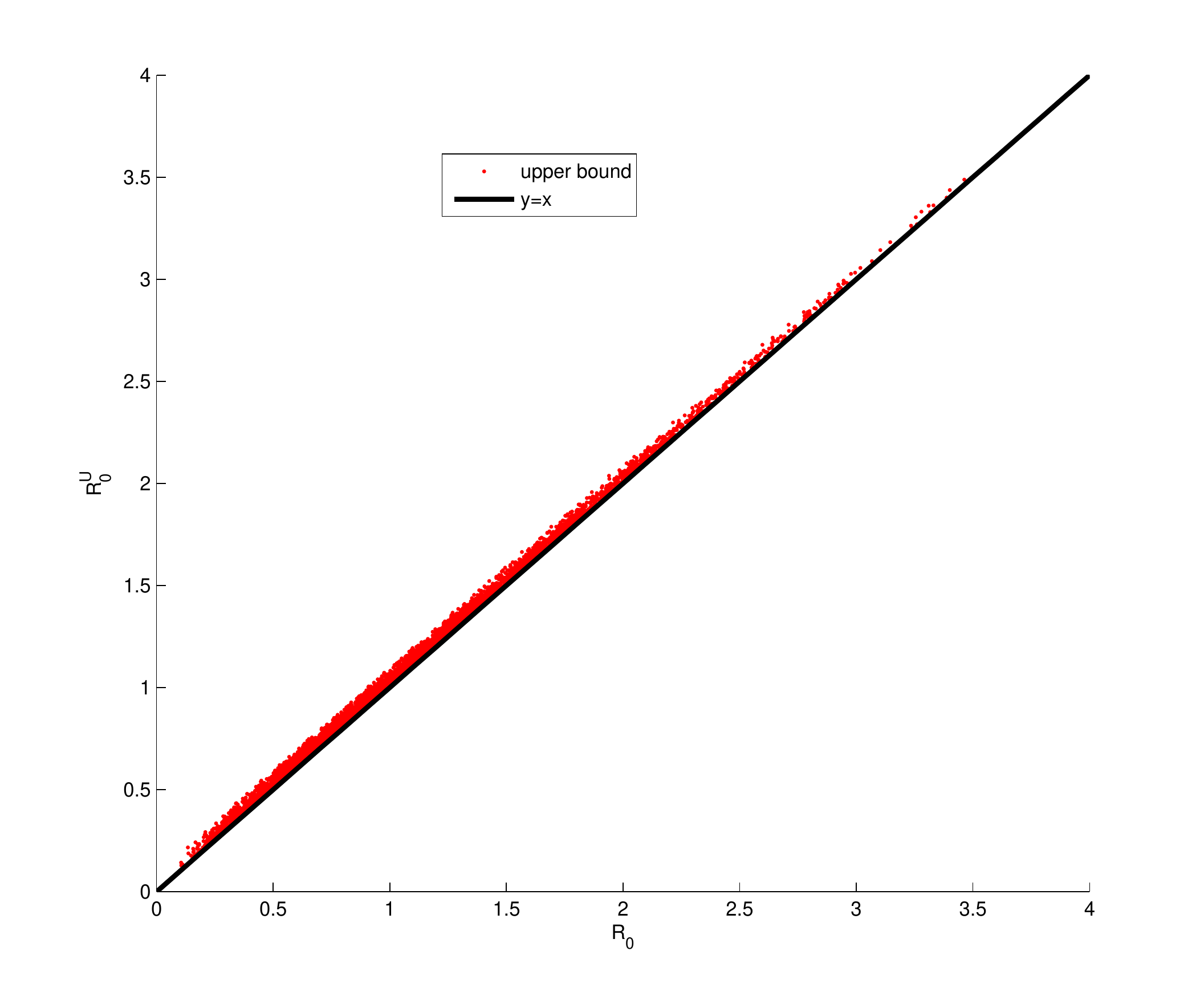}}
\hspace{0.1in}
\end{figure}

\begin{figure}[!htbp]
\centering
\subfigure[The reproduction number and its lower  bound.]{
\label{fig:lowerbound}
\includegraphics[angle=0,width=10.9cm,height=10.9cm]{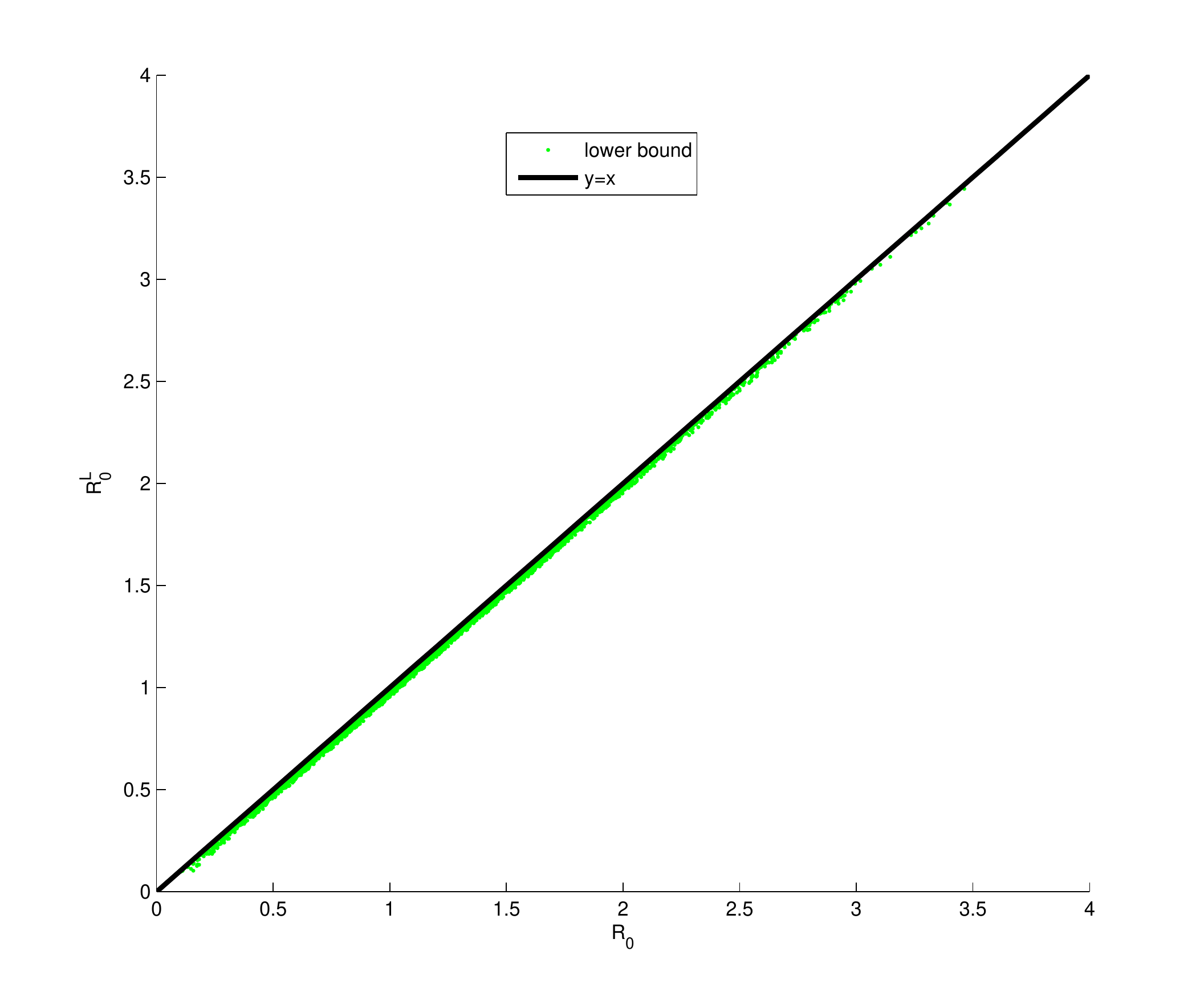}}
\hspace{0.1in}
\caption{The reproduction number and its upper and lower bound}
\label{fig:ApproximateR0}
\end{figure}

$R_0^H$ and   $R_0^H +q_1 $ are the lower bound and the upper bound of $R_0$, respectively. Therefore, $R_0^H +q_1 <1 \Rightarrow  R_0<1$, and $R_0^H >1 \Rightarrow R_0 >1$.  To verify that the derived bounds are tight, we perform extensive simulations using $5000$ sets of parameters uniformly distributed within the range defined in \cite{Gaff2007}. In Figure   \ref {fig:upperbound} and  \ref {fig:lowerbound}, $R_0^U=R_0^H+q_1$ vs. $R_0$ and  $R_0^L=R_0^H$ vs. $R_0$ are plotted, respectively.
First, the difference between the exact values and each bound is very small. In fact, the red and green points lay very close to the line $y=x$. Additionally, in the case of the upper bound, the red points are just slightly above the line $y=x$, while in the case of the lower bound, the green points are just slightly below the line $y=x$.\\

\subsection*{ Biological Interpretation of Bounds for $R_0$}
The bounds for $R_0$, as given in inequalities (\ref{equation:R0boundd}), can be interpreted biologically as follows.
The lower bound, $R_0^H$,  is the reproduction number for horizontal transmission because $R_0^H=\rho\ (F_HV_H^{-1})$, where $\rho\ (F_HV_H^{-1})$ represents the spectral radius of the next generation matrix for horizontal transmission  $F_HV_H^{-1}$.  The upper bound is given by the sum of $R_0^H$ and a second term that is only related to vertical transmission, i.e. from mothers to their offspring in the \it Aedes \rm mosquito population.

 $R_0^H$  includes \it Aedes\rm-livestock  interaction and   \it Culex\rm-livestock interaction. More specifically, we define the reproduction number due to the interaction between \it Aedes \rm and livestock represented by $R_0^{H(A-L)}$ as  $$R_0^{H(A-L)}=\sqrt{
\frac{\varepsilon_2}{(b_2+\varepsilon_2)
(b_2+\gamma_2+\mu_2)}
\frac{\varepsilon_1\beta_{12}\beta_{21}}{b_1
 (b_1+\varepsilon_1)}}$$

$R_0^{H(A-L)}$ can be rewritten as follows.\\
\begin{align}
R_0^{H(A-L)}&=\sqrt{
\Big[ \frac{\beta_{12}} {d_1\frac{N_1^*}{K_1}}
\frac{\varepsilon_1}{(d_1\frac{N_1^*}{K_1}+\varepsilon_1)}\Big]
\Big[ \frac{\beta_{21}} {(d_2\frac{N_2^*}{K_2}+\gamma_2+\mu_2) }\frac{\varepsilon_2}{(d_2\frac{N_2^*}{K_2}+\varepsilon_2)
}\Big] }\label{equation:R0AL}
\end{align}
because
\begin{align*}
b_1 &=d_1 \frac{N_1^*}{K_1}\\
b_2 &=d_2\frac{N_2^*}{K_2}\\
\end{align*}
where:\\
$N_1^*=$ the total number of \it Aedes \rm mosquitoes at disease free equilibrium.\\
$N_2^*=$ the total number of livestock at disease free equilibrium.\\

$R_0^{H(A-L)}$ consists of the product of four terms.  Each infected \it Aedes \rm  \rm mosquito  can infect $\frac{\beta_{12}}{d_1 \frac{N_1^*}{K_1}}$  susceptible livestock throughout its lifetime.  Similarly, each infected livestock can infect   $\frac{\beta_{21}}{d_2\frac{N_2^*}{K_2}+\gamma_2+\mu_2}$  susceptible \it Aedes \rm mosquitoes during its lifetime. The probability of \it Aedes \rm  mosquitoes and livestock surviving through the incubation period to the point where  they become infectious is $\frac{\varepsilon_1}{d_1 \frac{N_1^*}{K_1}+\varepsilon_1} $ and $\frac{\varepsilon_2}{d_2\frac{N_2^*}{K_2}+\varepsilon_2} $, respectively. Therefore, $R_0^{H(A-L)}$  is  the geometric mean of the average number of  secondary  livestock infections produced by one \it Aedes \rm mosquito vector in the first square bracket in (\ref{equation:R0AL}), and the average number of secondary \it Aedes \rm mosquito vector infections  produced by one livestock host in the second square bracket  in (\ref{equation:R0AL}). \\

Similarly,  we define the reproduction number due to the interaction between \it Culex \rm and livestock represented by $R_0^{H(C-L)}$  as $$R_0^{H(C-L)}=\sqrt{\frac{\varepsilon_2}{(b_2+\varepsilon_2)
(b_2+\gamma_2+\mu_2)}\frac{\varepsilon_3\beta_{32}\beta_{23}}{b_3(b_3+\varepsilon_3)}}$$

We can rewrite $R_0^{H(C-L)}$ as follows.\\
\begin{align}
R_0^{H(C-L)}&=\sqrt{\Big[\frac{\beta_{32}}{ d_3\frac{N_3^*}{K_3} }\frac{\varepsilon_3}{(d_3\frac{N_3^*}{K_3}+\varepsilon_3)} \Big]\Big[\frac{\beta_{23}}{  (d_2\frac{N_2^*}{K_2}+\gamma_2+\mu_2)}\frac{\varepsilon_2}{(d_2\frac{N_2^*}{K_2}+\varepsilon_2)}\Big]} \label{equation:R0CL}
\end{align}
because
\begin{align*}
b_3 &=d_3 \frac{N_3^*}{K_3}
\end{align*}
where:\\
$N_3^*=$ the total number of \it Culex \rm mosquitoes at disease free equilibrium.\\

 $R_0^{H(C-L)}$ also consists of the product of four terms.  Each infected  \it Culex \rm mosquito can infect $\frac{\beta_{32}}{d_3 \frac{N_3^*}{K_3}}$ susceptible livestock throughout its lifetime.  Similarly, each infected livestock can infect  $\frac{\beta_{23}}{d_2\frac{N_2^*}{K_2}+\gamma_2+\mu_2}$    susceptible \it Culex \rm mosquitoes. The probability of \it Culex \rm mosquitoes surviving through the incubation period to the point where  they become infectious is  $\frac{\varepsilon_3}{d_3 \frac{N_3^*}{K_3}+\varepsilon_3} $. Similarly,  the probability of livestock surviving through the incubation period to the point where  they become infectious is $\frac{\varepsilon_2}{d_2\frac{N_2^*}{K_2}+\varepsilon_2} $.
Therefore, $R_0^{H(C-L)}$ is the geometric mean of the  average number of the secondary livestock infections  produced by one \it Culex \rm mosquito vector in the first square bracket  in (\ref{equation:R0CL}), and the average number of secondary \it Culex \rm mosquito vector infections  produced by one livestock in the second square bracket  in (\ref{equation:R0CL}).

\begin{figure}[h]
\centering
\includegraphics[angle=0,width=5cm,height=5cm]{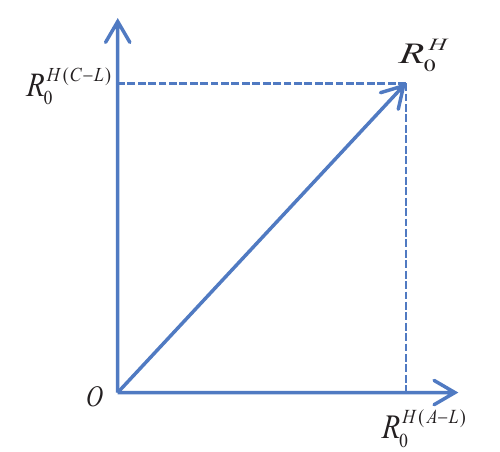}
\hspace{0.1in}
\caption{The interpretation of $R_0^H$}
\label{fig:R0H}
\end{figure}

 The expression (\ref{equation:R0H1}) for $R_0^H$, can be rewritten as  $R_0^H=\sqrt{({R_0^{H(A-L)}})^2+({R_0^{H(C-L)}})^2}$ , where the dependence of  $R_0^H$ on $R_0^{H(A-L)}$ and $R_0^{H(C-L)}$ is shown in  Figure \ref{fig:R0H}. The  square root is due to the vector-host-vector viral transmission path \cite{ Gaff2007,  heffernan2005perspectives, Deikmann1995}.   Obviously, the horizontal reproduction number  increases with the increase of each of the four terms in $R_0^{H(A-L)}$ and $R_0^{H(C-L)}$.\\

\bibliographystyle{model2-names}
\bibliography{xuelzbExport}

\end{document}